\documentclass[fleqn,usenatbib]{mnras}

\usepackage{newtxtext,newtxmath}
\usepackage{xcolor}
\usepackage[dvipsnames]{xcolor}
\usepackage{booktabs, caption, subcaption} 
\usepackage{chngcntr}

\DeclareRobustCommand{\VAN}[3]{#2}
\DeclareUnicodeCharacter{2212}{-}
\let\VANthebibliography\thebibliography
\def\thebibliography{\DeclareRobustCommand{\VAN}[3]{##3}\VANthebibliography}

\usepackage[T1]{fontenc}
\usepackage{enumerate}
\usepackage{booktabs}
\usepackage{multirow}

\usepackage{graphicx, float}	
\usepackage{amsmath}	

\title[Anisotropic wind in TDEs]{Anisotropic wind in tidal disruption events}

\author[P. Martire et al.]{
P. Martire$^{1}$\thanks{E-mail: martire@strw.leidenuniv.nl}, 
E.M. Rossi,$^{1}$
N. C. Stone,$^{2,3}$
E.Steinberg,$^{3}$
I.Giron$^{3}$
\\
$^{1}$Leiden Observatory, Leiden University, PO Box 9513, 2300 RA Leiden, The Netherlands \\
$^{2}$Department of Astronomy, University of Wisconsin, Madison, WI 53706, USA\\
$^{3}$Racah Institute of Physics, The Hebrew University, Jerusalem, 91904, Israel\\
}
\date{Accepted XXX. Received YYY; in original form ZZZ}

\pubyear{\the\year{}}

\begin{document}
\pagerange{\pageref{firstpage}--\pageref{lastpage}}
\maketitle

\begin{abstract}
Over the coming years, the number of tidal disruption events (TDEs) is expected to substantially increase with observations from the Vera Rubin Observatory (g and r band) and {\it ULTRASAT} (near UV) wide-field surveys.  
These future samples have great promise to characterize the bottom end of the massive black hole mass function, but existing detections of intermediate mass black hole TDEs are primarily in X-rays, leaving their optical/UV emission largely unexplored.
We present a time- and angle-dependent analysis of the outflow produced by dissipation near pericentre in a three-dimensional end-to-end radiation-hydrodynamics simulation of a TDE by a $10^4 M_\odot$ black hole with realistic parameters, run with the code \texttt{RICH}. We find that outflow anisotropy produces viewing-angle-dependent observables. Towards the poles and the pericentre region,  mass-loss rates are low and bolometric luminosities reach $\sim2$--$3$ times the Eddington luminosity. Towards the stream, the properties show a stronger dependence on latitude: the mass-loss rate increases and the bolometric luminosity decreases as the line of sight approaches the orbital plane. These denser regions favour H$\alpha$ and H$\beta$ emission. Despite these variations, all viewing directions show a common spectral evolution, with an initial soft X-ray flare followed (around $1.25t_{\rm fb}\approx3$~days) by the reprocessing of shock-powered emission into the UV and optical bands.  Although the optical/UV luminosities we predict for this TDE are likely too dim for past surveys (e.g. ASAS-SN, ZTF), they are within the detection capabilities of LSST and {\it ULTRASAT} to horizons of $\sim 790$ and $\sim 340$ Mpc, respectively, for the brightest viewing directions.
\end{abstract}



\section{Introduction}
\begin{figure*}
\centering
    \includegraphics[width=\linewidth]{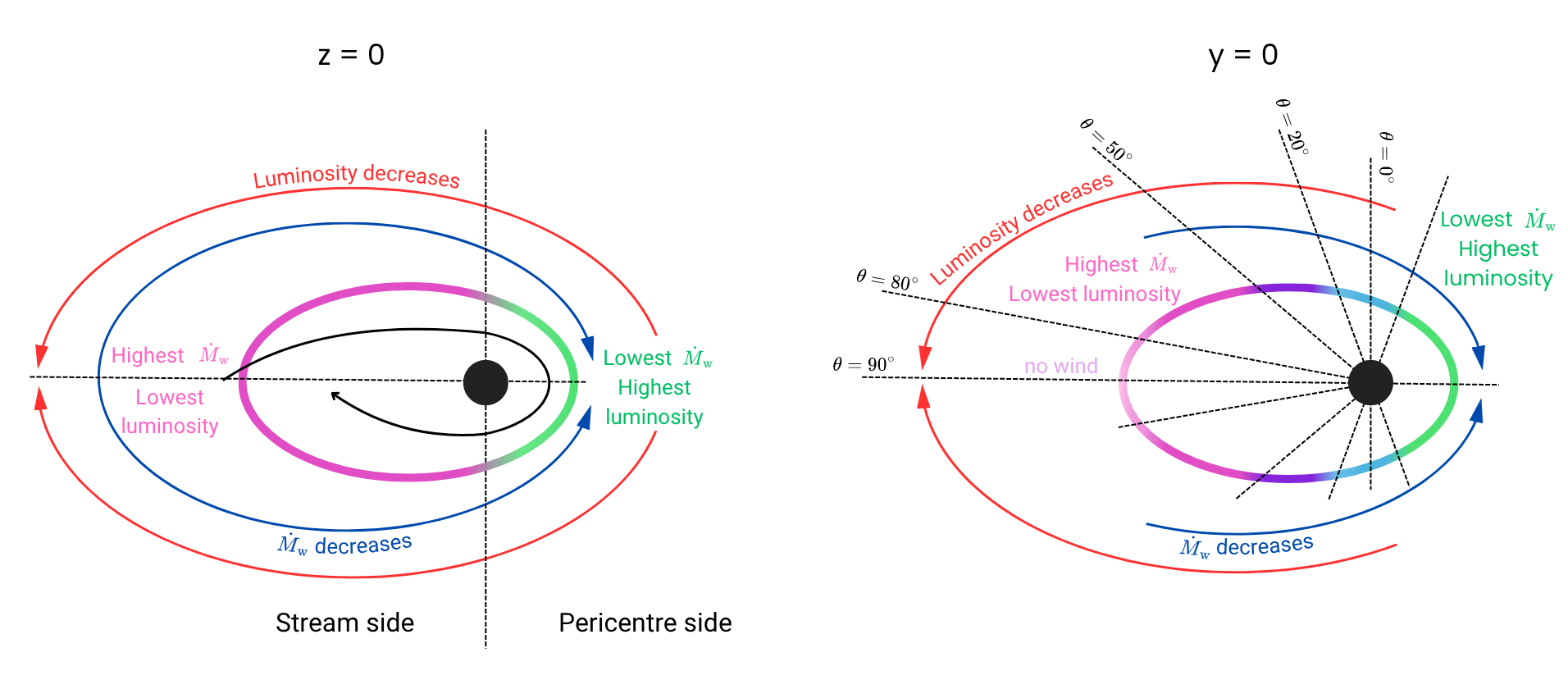}
    \caption{Schematic drawing illustrating the angular geometry of the wind mass-loss rate $\dot{M}_{\rm w}$ and the observed luminosity at the peak of the bolometric light curve ($L_{\rm FLD}$). An anisotropic radiation-dominated wind is launched from the pericenter region and expands away from the black hole. While the outflow is approximately symmetric with respect to the orbital plane, its properties depend on both the polar and azimuthal viewing angle. 
    We identify the following major regions: the poles ($0^\circ\leq\theta<20^\circ$; $160^\circ < \theta \le 180^\circ$), the {\it pericentre side} ($20^\circ\leq\theta<160^\circ$ and $x\geq0$), and the {\it stream side} ($20^\circ\leq\theta<160^\circ$ and $x<0$). The latter is further divided in latitude according to the following boundaries (which are indicated in the right panel): $20^\circ\leq\theta<50^\circ, \,50^\circ\leq\theta<80^\circ,$ and $80^\circ\leq\theta\leq90^\circ$.
    The highest luminosities are observed along the polar directions, with similarly high values in the pericentre side (where there is little variation with latitude).  In contrast, the stream side is dimmer, with luminosity decreasing towards the orbital plane. The angular trend in $\dot{M}_{\rm w}$ is reversed, reaching its maximum in the stream side and its minimum towards the poles and pericentre side.
    \emph{Left panel}: orbital ($z=0$) plane. \emph{Right panel}: $y=0$ plane.} 
\label{fig:sketch}
\end{figure*}
When a star gets close enough to a Massive Black Hole (MBH) that tidal forces exceed its self-gravity, it is disrupted in what is known as a Tidal Disruption Event (TDE), producing a bright flare. These transients are essential tools for MBH demographics, allowing us to study the formation and growth of quiescent MBHs. Moreover, they can be used to probe stellar dynamics \citep{Stone20, Shepherd26}, host galaxy properties \citep{LawSmith17, French20}, and accretion \citep{Cannizzo90, Shen14, Mummery19, Alush25, Mummery25, Curd26} and jet physics \citep{Stone12, Tchekhovskoy14, Coughlin20, Dai21, TeboulMetzger23}.
TDEs from Intermediate-Mass Black  Holes (IMBHs, i.e., MBH with masses $M_{\rm BH}\in [10^4-10^6]M_\odot$) are of particular interest, as their detection and characterisation would prove the presence of these elusive mid-weight compact objects, and would provide an observational bench for testing MBH formation models \citep{Greene20, Patra25, Li25, Wang26}. 

Currently, there are $\sim 100$ observed TDEs, spanning a wide range of wavelengths across the electromagnetic spectrum, from radio \citep{Dykaar24, Goodwin25}, through mid-infrared \citep{Mattila18, Masterson24}, to X-ray \citep{Saxton20, Sazonov21, Grotova25}. The majority, however, have been detected in the UV and optical bands \citep{Gezari08, vanVelzen11, vanVelzen20, vanVelzen21, Yao23}, and their number is expected to increase by 1-2 orders of magnitude with upcoming data from the Rubin/LSST survey \citep{BricmanGomboc20}. Understanding the origin of the early-time emission -- the brightest, most easily observed, and most hotly debated emission component -- 
and its connection to underlying event parameters is therefore of extreme importance to maximise the scientific return of this wealth of upcoming data.

It was once assumed that the stellar debris would quickly settle in a disc around the MBH, with the observed light curve being powered by accretion \citep{Rees88, Strubbe09, LodatoRossi11}.
While in agreement with {\it late-time} optical and UV observations \citep{vanVelzen19}, this model has been challenged by discrepancies with the {\it early-time} optical emission properties \citep{vanVelzen11, Gezari12, Arcavi14, Hung17, Nicholl19}. These include high optical luminosities ($\sim10^{43}-10^{45}$erg/s), modest effective temperatures ($\approx 2-5 \times 10^{4}$ K), large photosphere radii ($\approx 10^{14}-10^{15}$ cm), and spectral line widths implying gas velocities of $10^{4}$ km/s, all of which are difficult to reconcile with expectations from compact accretion discs. 
Alternative mechanisms have been proposed for powering the early-time optical emission of TDEs, either through the dissipation of orbital energy in shocks embedded in optically thick debris \citep{Piran15, Shiokawa15, Bonnerot17, Svirski17, Ryu20}, or through reprocessing of high-energy radiation from an inner source by an extended, optically thick layer. In the former scenario, shock energy is thermalised locally and emerges as optical/UV radiation, whereas in the latter the reprocessing material is spatially distinct from the primary high-energy source\footnote{Hybrid scenarios are also possible, in which radiation generated by shocks, possibly together with radiation from an accretion flow, is further reprocessed by an extended photosphere.}. The origin and structure of such an extended reprocessing layer remain uncertain.
Associated to either an accretion disc or shocks, it has been described variously as an Eddington envelope \citep{Loeb97, Price24}, a quasi-static or cooling TDE envelope \citep{CoughlinBegelman14, Roth16, Metzger22, Tuna26}, a mass-loaded outflow \citep{MetzgerStone16}, a collision-induced outflow \citep{LuBonnerot20}, or a super-Eddington wind \citep{Strubbe09, LodatoRossi11,  Miller15, Martire26, Mockler26}.  Many of these qualitative descriptions overlap significantly with each other, though one clear distinction is between unbound, outflowing reprocessing layers and bound, quasi-static ones. In this framework, early- or late-time X-ray detections can be interpreted as phases during which the reprocessing layer is absent because, respectively, it has not yet formed or it has dissipated. 
However, the simultaneous detection of both X-ray and optical/UV luminosity  \citep{Holoien16b, Holoien16, Homan23, Cao24, Guolo24, Malyali26} remains challenging to reconcile with such models.

A unified scheme has been proposed \citep{MetzgerStone16, Roth16, Dai18, Thomsen22, Parkinson25} to account for both X-ray and optical emission as function of the system geometry and the observer viewing angle. In this scenario, X-ray photons escape preferentially through low-density polar funnels, while at higher latitudes (i.e., towards the mid-plane) they are intercepted and reprocessed into optical and ultraviolet radiation. 
The time-dependent evolution of the outflow may further shape the observed emission \citep{Mockler26}. 
However, predictions for emission reprocessing have often relied on simulations that do not follow the full evolution of the system end-to-end or that assume one-dimensional geometries, potentially overestimating photon trapping and reprocessing efficiency \citep{Parkinson25}.
More recently, \citet{Parkinson25} relaxed this assumption by performing Monte Carlo radiative transfer on a non-spherical outflow, highlighting the importance of fully three-dimensional (3D) simulations for understanding the radiative transfer and the observed emission.\\

In this work, we analyse the outflow produced in an end-to-end 3D radiation-hydrodynamics simulation of a main sequence star disrupted by an IMBH. We first provide a qualitative overview of the results in Fig.\ref{fig:sketch}. The simulation setup and methodology are detailed in Sec.\ref{sec:meth}, followed by a quantitative analysis in Sec.\ref{sec: res}.  We discuss our results in Sec.\ref{sec: disc}, and we draw conclusions in Sec.\ref{sec: concl}.

\section{Methods}
\label{sec:meth}
We present further analysis of the simulation first described in \citet{Martire26}, which followed the 3D disruption of a star (with radius $r_\star = 0.47R_\odot$, mass $m_\star=0.5M_\odot$ and polytropic index $n=1.5$) by a non-spinning BH with mass $M_\text{BH}=10^4M_\odot$. The BH is placed at the origin of the reference frame and the star is initially positioned at $3r_\text{t}$ on a parabolic orbit with pericentre $r_\text{p}$ at $(x,y,z)=(r_\text{t}, 0,0)$, where $r_{\rm t}= r_\star(M_\text{BH}/M_\star)^{1/3}\approx13R_\odot\approx600r_{\rm g}$ is the tidal radius and $r_{\rm g}=GM_{\rm BH}/c^2$ is the gravitational radius (with $G, c$ being the gravitational constant and the speed of light, respectively). 
We evolved the system through the radiation-hydrodynamic code \texttt{RICH}\footnote{\url{https://gitlab.com/eladtan/RICH/-/tree/master?ref_type=heads}} \citep{rich} with a Paczyński–Wiita potential \citep{pseudopot} in order to approximate the leading-order effect of relativistic apsidal precession. Additional details of the numerical implementation can be found in \citet{Martire26}; we include here only the aspects relevant to this study. 
We use a realistic equation of state \citep{Tomida13} and treated radiation according to a flux limited diffusion (FLD) scheme under the approximation of grey transport, considering both absorption and scattering opacity. We followed \citet{Krief16} and used tabulated values for opacities,
assuming solar abundances and local thermodynamic equilibrium. Absorption, free-free, bound-free and bound-bound interactions are considered and we used frequency-mean opacities (see Appendix \ref{app: opac}). We adopted an initial number of $N\approx4\times10^6$ cells and run the simulation up to $t\approx2.2t_{\rm fb}$, where $t_{\rm fb}= 4 \,\text{days}\,(M_\text{BH}/(10^4M_\odot))^{1/2}(M_\star/M_\odot)^{-1} (r_\star/R_\odot)^{3/2}\approx 2.5 \,\text{days}$ is the approximate fallback time  of the most bound debris \citep{Rees88}, after which $N\approx 7\times 10^7$.

After disruption, the stellar debris is imparted with a spread in orbital energy $\Delta\varepsilon=GM_{\rm BH}r_\star/r_{\rm t}^2$ \citep[see][for a review of the stellar disruption stage]{Rossi21}. Approximately half of the star gets ejected to infinity, while
the rest returns to the BH, eventually circularizing in an accretion disc. 
Past work \citep{Blandford99, Ohsuga11, Jiang14} has shown that outflows are produced in axisymmetric discs with super-Eddington accretion rates. While early stages of TDEs may not be axisymmetric, initial mass fallback rates will be super-Eddington for full disruptions of main sequence stars provided $M_{\rm BH} \lesssim {\rm few} \times 10^6 M_\odot$, suggesting the likely importance of outflows.  
Indeed, \citet{Martire26} showed that, when the stellar debris returns to the black hole after disruption, it is compressed in the vertical direction during its pericentre passage through the \emph{nozzle shock} and a radiation-dominated, optically thick wind forms here due to the super-Eddington nature of the dissipation rate. 

In this paper, we further analyse the properties of this wind. We adopt spherical coordinates $(r, \theta, \varphi)$ and divide the domain into distinct regions (see, Fig.\ref{fig:sketch}). Due to symmetry about the orbital plane, we will refer only to the northern hemisphere ($0^\circ\leq\theta\leq90^\circ$).
The \emph{north pole} is defined by $0^\circ\leq\theta<20^\circ$, and the remaining $\theta$ range is additionally divided into two sectors based on the sign of the Cartesian $x$-coordinate: the \emph{pericentre side} ($x\geq0$),  containing the pericentre direction, and the \emph{stream side} ($x<0$) containing the debris stream. The latter is further split in three regions due to its high degree of anisotropy: $20^\circ\leq\theta<50^\circ$ (i.e., next to the pole, from now on called {\it high stream side}), $50^\circ\leq\theta<80^\circ$ ({\it middle stream side}), and $80^\circ\leq\theta\leq90^\circ$ (i.e., next to the orbital plane, thereafter {\it eccentric flow side}).

\begin{figure*}
\centering
    \includegraphics[width=\linewidth]{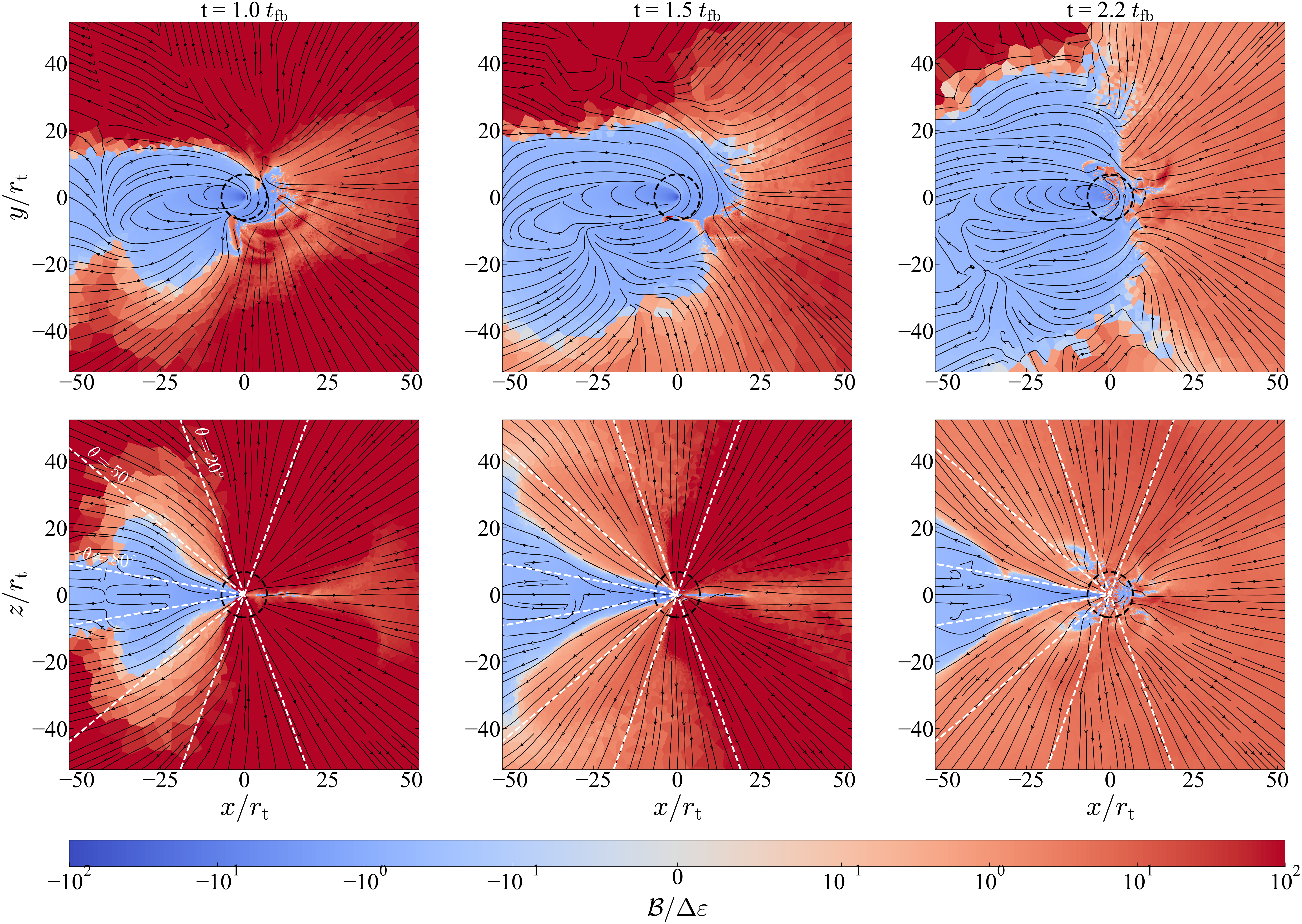}
    \caption{Wind evolution as seen in the orbital, $z=0$ plane (\emph{first row}) and from the $y=0$ plane (\emph{second row}). Each column shows a different snapshot in time.  Points are colour-coded with their Bernoulli parameter $\mathcal{B}$ (normalized by the dynamical energy spread $\Delta\varepsilon$). White dashed lines in the $xz$ plane show the subdivision of the space in angular regions described in Sec.\ref{sec:meth}. The dashed black circle is located at $r=0.5a_{\rm mb}$, where the outflow/wind rates of Fig.\ref{fig:Mdot_M} are computed. The black flow vectors show that an asymmetrical wind is expanding from the orbital plane to lower latitudes, becoming more spherical and following increasingly radial trajectories over time.} 
\label{fig:bern}
\end{figure*}
For each sector, we calculate characteristic radii, mass loss rates, luminosities and spectra as described below.
Due to the different surface areas of the sectors, we compute the isotropic-equivalent mass loss rates, $\dot{M}$, and advection luminosity, $L_{\rm adv}$, at a given radius $\bar{r}$ as 
\begin{align}
    \dot{M}_{\rm iso}(t)&= C(t)\dot{M}(t)= C(t)\sum_{i:r_i= \bar{r}} \mathcal{A}_i(t)\rho_i(t) v_{{\rm r},i}(t), \label{eq: Mdot iso}\\
    L_{\rm adv, iso}(t) &= C(t)L_{\rm adv}(t) = C(t)\sum_{i:r_i= \bar{r}} \mathcal{A}_i(t) u_{{\rm rad},i}(t) v_{{\rm r},i}(t) \label{eq:Ladv iso},
\end{align}
where $\rho$ is the mass density, $v_{\rm r}$ the radial velocity, $u_{\rm rad}$ the radiation energy density, $\mathcal{A}$ the surface area of a cell, $C(t)\equiv 4\pi \bar{r}^2/(\sum_i  \mathcal{A}_i(t))$, and the summation is performed over cells located in the sector under consideration. 
These isotropic-equivalent quantities can be understood as the values of mass loss or luminosity that would be imputed by an external observer only able to measure properties of a single viewing angle, but who then extrapolated to $4\pi$ steradians assuming isotropy.  When computing polar profiles (see, Sec.\ref{sec: res}), we use Eqs.\eqref{eq: Mdot iso}, \eqref{eq:Ladv iso}, but summing over azimuthal angles of cells that are at the same latitude in the same sector, at a fixed radius.\\

We post-process single-band light curves and spectral energy distributions (SEDs) assuming black-body radiation from each simulation cell. First we find the thermalization surface, i.e., the outermost surface at which photons, in local thermal equilibrium with matter, can scatter without being reabsorbed.
We discretize the 3D space with \texttt{HEALPix} \citep{healpix}, creating $N_\text{obs}=192$ observers spacing the same solid angle. Along each direction, we trace a logarithmically spaced radial ray and define the thermalization radius as the radius $r_{\rm th}$ such that $\tau_{\rm eff} (r_{\rm th})=1$, where 
\begin{equation}
    \label{eq:colorsphere}
    \tau_{\rm eff} (r) \equiv \int_{r}^\infty  \sqrt{3\alpha_{\rm abs}(\alpha_{\rm abs} + \sigma_{\rm s})} \rm{d}r',
\end{equation}
is the effective optical depth, with $\alpha_{\rm abs}$ (units of cm$^{-1}$) being the absorption coefficient, and $\sigma_{\rm s}$ (units of cm$^{-1}$) the scattering coefficient (see Appendix \ref{app: opac} for further details). We compute the luminosity emitted from each cell along the ray as
\begin{equation}
    \tilde{L}_{\nu, \rm {cell}} =   \alpha_{\rm abs, \,cell}V_{\rm cell} e^{-\tau_{\rm eff, \,cell}} B_\nu(T_{\rm cell}),
\end{equation}
where $B_\nu$ is the Planck function, $T_{\rm cell}$ the temperature of the cell, and $V_{\rm cell} \equiv (4\pi/N_{\rm obs}) r_{\rm cell}^2 \rm{d}r$, with d$r$ being the radial width of the bin along the ray. 
The emission $\tilde{L}_\nu$ along an observer's line of sight is obtained by summing the contributions from all cells along that direction, from infinity down to a cutoff interior of the thermalization radius\footnote{We numerically perform the integration in Eq.\eqref{eq:colorsphere} radially inwards to $\tau_{\rm eff}=5$ to ensure we are well inside the thermalization surface.}.
To be consistent with the results in \citet{Martire26}, for each $i$-th observer we normalize as
\begin{equation}
\label{eq: Lnu}
    L_{\nu,i} = \frac{L_{{\rm FLD}}(r_{{\rm ph}, i})\tilde{L}_{\nu,i}}{\int \tilde{L}_{\nu,i}d\nu},
\end{equation}
where $L_{{\rm FLD}}(r_{{\rm ph}})$ is the flux-limited luminosity at the photosphere\footnote{We define the photosphere as the surface identified by $r_{\rm ph}$, where $\tau(r_{\rm ph}) \equiv \int^\infty_{r_\text{ph}}\alpha_\text{Ross}(r) {\rm d}r= 2/3$ and $\alpha_\text{Ross}$ is the Rosseland mean opacity coefficient.} as computed in \citet{Martire26} (Eq.11). 
To estimate the luminosity measured by an observer $i$, we account for photon scattering and the 3D geometry of the system by weighting the contribution of each  cell by its angular offset $\delta$ relative to the observer $i$, applying a factor $\max(0,\cos \delta)$.
Thus, for the $i$-th observer:
\begin{equation}
\label{eq:Lnu_weigh}
    \frac{\sum_{k=0}^{N_{\rm obs}}\max(0,\cos\delta_{k,i}) L_{\nu,k}}{\sum_{k=0}^{N_{\rm obs}}\max(0,\cos\delta_{k,i})}.
\end{equation}

In Appendix \ref{app: MG}, we further compare our results with those obtained self-consistently in the simulation presented in \citet{Giron26}, who evolve the same event described in our work (albeit at lower resolution) but under a multigroup diffusion approximation.\\
\begin{figure*}
    \centering
    \includegraphics[width=\linewidth]{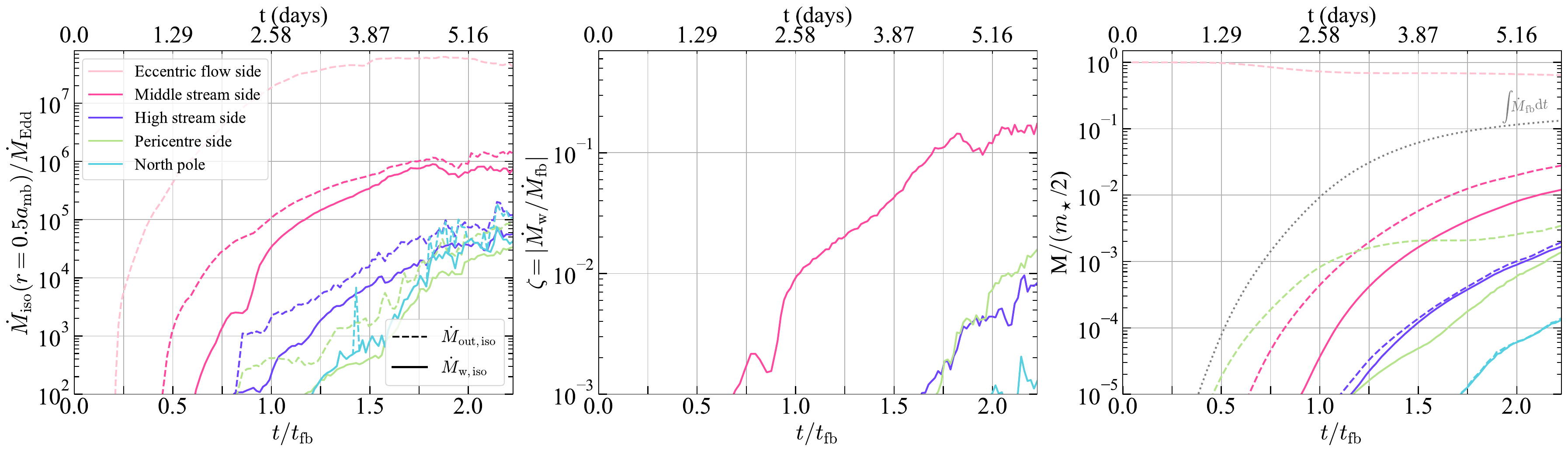}
    \caption{\emph{Left panel}: 
    Time evolution of the isotropic-equivalent mass loss rates for the outflow ($v_{\rm r}>0$ and $\mathcal{B}\gtrless 0$) and the wind ($v_{\rm r}>0$ and $\mathcal{B}>0$): $\dot{M}_{\rm out, iso}$ (dashed line) and $\dot{M}_{\rm w, iso}$ (solid line), respectively. The curves are computed at $r=0.5a_{\rm mb}$ as a $\{\theta,\varphi\}$- average on the spherical sectors (see the legend for the corresponding colours). \emph{Middle panel}: wind launching efficiency $\zeta \equiv |\dot{M}_{\rm w}/\dot{M}_{\rm fb}|$.
    \emph{Right panel}: total gas mass in the outflow (dashed line) and in the wind (solid line), normalized by the dynamically bound mass ($m_\star/2$). The dotted line represent the returned mass, i.e. $\int\dot{M}_{\rm fb}{\rm d}t$.
    Overall, more material becomes unbound and the mass ejection rates increase in time, remaining consistently super-Eddington. Most of the total unbound mass leaves in the middle stream side, which by the end of the simulation transports out the 10\% of the returning stream material (with $ \zeta\approx0.2$).  
    }
    \label{fig:Mdot_M}
\end{figure*}

Finally, we compute the ionization state and line emissivities at the photosphere applying the approximate analytic framework of \citet{Aspegren26} to each of the $N_{\rm obs}$ observers.
Assuming local thermodynamic equilibrium, we compute the specific continuum luminosity as
\begin{equation}
\label{eq:Lcont}
    L_{\lambda, {\rm cont}} = 4\pi^2 r^2_{\rm ph}B_{\lambda}(T_{\rm ph}).
\end{equation}
For the specific line luminosity, we iteratively solve the  coupled Saha equations for a gas cloud composed of hydrogen and helium. 
The ratio between the number densities $n_i$ and $n_{i+1}$ of two consecutive ionization stages with partition functions $Z_i$ and $Z_{i+1}$ is given by
\begin{equation}
    \frac{n_{i+1}}{n_i}=\Omega, \quad \text{where} \,\,\Omega= \frac{2}{n_{\rm e}\lambda^3_{\rm T}}\frac{Z_{i+1}}{Z_i}\exp\left(-\frac{\chi}{k_{\rm B}T}\right),
\end{equation}
where $n_{\rm e}$ is the number density of free electron, $\lambda_{\rm T}\equiv h/\sqrt{2\pi m_{\rm e}k_{\rm B}T}$ the electron thermal deBroglie wavelength, $\chi$ the ionization potential, and $k_{\rm B}$ the Boltzmann constant. The partition function of an ionization stage is defined as
\begin{equation}
    Z(T)=\sum^{n_{\rm max}}_{n=1}g_ n\exp\left(-\frac{\chi_{n}}{k_{\rm B}T}\right),
\end{equation}
where $g_n$ and $\chi_n$ are the statistical weight of state $n$ (which takes into account the degenerate sublevels) and its energy level relative to the ground state.
Considering Boltzmann distribution for the population of excited states above the ground state, the line strength for the transition from the level $u$ can be expressed approximately \citep{Aspegren26} as
\begin{equation}
\label{eq:Lline}
    L_{\lambda, {\rm line}}(r_{\rm ph})\approx n_i(r_{\rm ph})\frac{g_u}{Z_i} \exp\left(-\frac{\chi_u}{k_{\rm B}T(r_{\rm ph})}\right)A\frac{hc}{\lambda}
    \frac{4\pi r^2_{\rm ph}\Delta r}{\Delta\lambda},
\end{equation}
where $g_u$ and $\chi_u$ denote the statistical weight and excitation energy of the level $u$ with respect to the ground state, and $h$ is the Planck constant. Finally, to determine the ratio $L_{\lambda, {\rm line}}/L_{\lambda, {\rm cont}}$ in each sector, we compute the median value of this ratio among the observers located within the sector.

We stress that this estimate of line luminosities is an approximate one, and precise calculations would likely require Monte Carlo radiation transfer post-processing with e.g. SEDONA \citep{Roth16, Roth18, Dai18, Thomsen22, Mockler26, Thomsen26}.  In particular, the assumption that most line emission comes from a frequency-averaged photosphere is crude, and neglects the stratified structure that sets zones of line emission in even spherical winds \citep{Roth16}.  Though a full Monte Carlo radiative transfer analysis is beyond the scope of this paper and must be deferred for future work, it will nevertheless be useful here to obtain even approximate estimates for line strengths, as optical emission (both continuum and line) in IMBHs is a largely unexplored frontier of TDE observation.

\section{Results}
\label{sec: res}
\begin{figure*}
    \centering
    \includegraphics[width=0.95\linewidth]{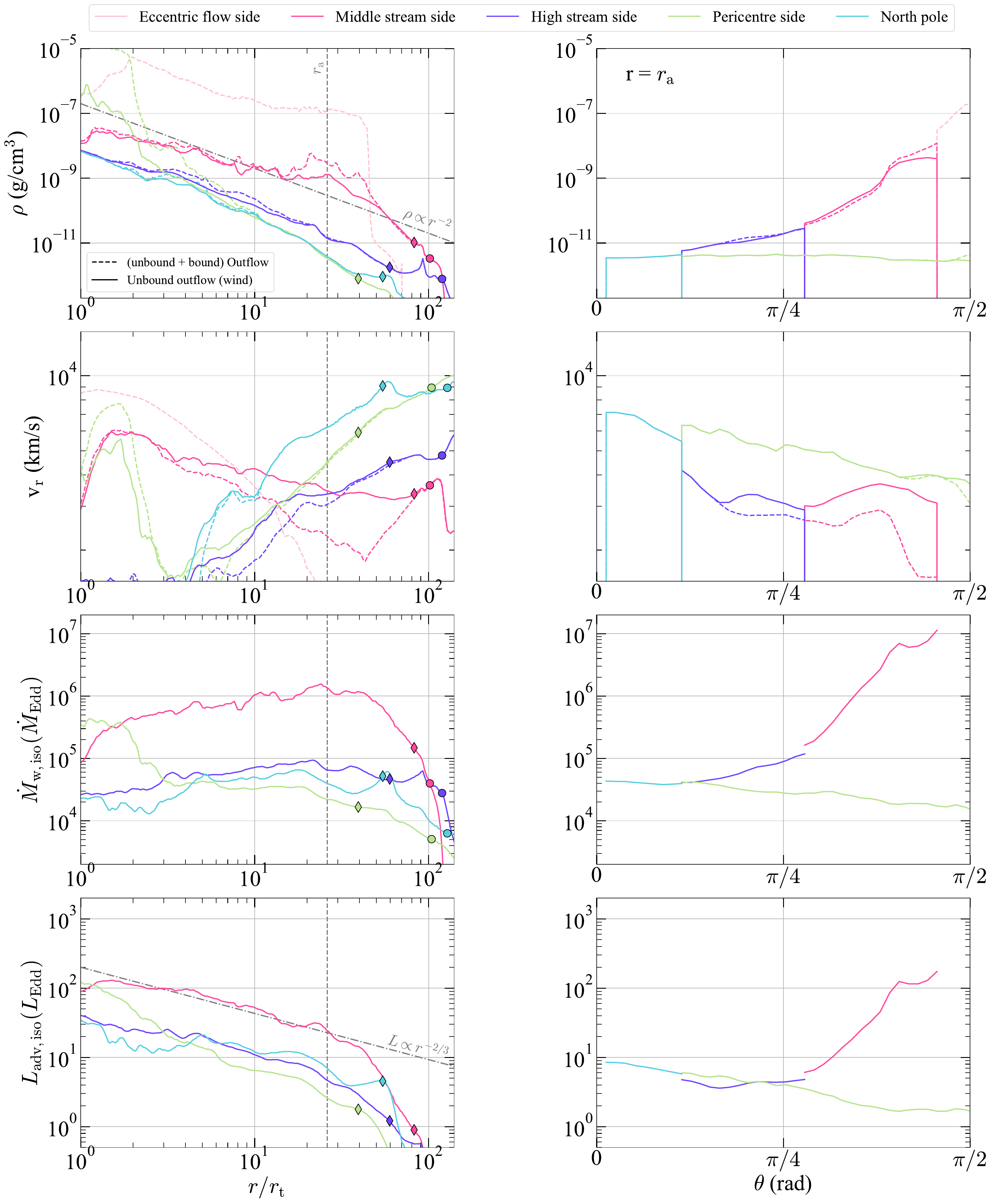}
    \caption{Radial (\emph{left column}; $\{\theta, \varphi \}$-averaged) and polar (\emph{right column}; $\varphi$-averaged) profiles at the end of the simulation ($t=2.2t_{\rm fb}$). \emph{First row}: mass-weighted gas density profiles. The dash-dotted line in the left panel shows the power-law $\rho\propto r^{-2}$ expected for steady winds. \emph{Second row}: mass-weighted radial velocity profiles. \emph{Third row}: isotropic-equivalent wind mass loss rate profiles. \emph{Fourth row}: isotropic-equivalent wind advection luminosity profiles. The dash-dotted line in the left panel shows the analytical $L\propto r^{-2/3}$ predicted for radiation-driven winds.
    Dashed-coloured lines are for all outflowing material (bound and unbound), while solid lines restrict to the wind (i.e., unbound outflowing). The vertical gray dashed lines in the left column denote the apocenter $r_{\rm a}$, which is where the polar profiles are computed. Diamonds and dots on the lines mark, respectively, the median $r_{\rm tr}$ and $r_{\rm ph}$ among the lines of sight in the selected region that contain an advective wind. The polar regions and the pericenter side show similar properties: the outflow is predominantly unbound, characterized by low densities and high velocities. The stream side near the orbital plane reaches higher densities, leading to greater mass outflow rates (though of primarily {\it bound} outflows). Overall, in the stream side, photons are advected over longer distances (i.e., larger $r_{\rm tr}$), resulting in a greater fraction of the energy being converted from radiation into kinetic energy.}
    \label{fig:prof}
\end{figure*}
In this section, we present the results of the analysis described above, performed by separating the previously defined regions (north pole, pericentre side, and the subdivided stream side) in order to capture and characterise the anisotropies of the wind.

Fig.\ref{fig:bern} illustrates the evolution of the stellar debris in the $z=0$ and $y=0$ planes, where points are colour-coded by their Bernoulli parameter $\mathcal{B}\equiv\varepsilon+\varepsilon_{\rm th}+(P+4u_{\rm rad}/3)/\rho$ , with $\varepsilon$ and $\varepsilon_{\rm th}$ being respectively the gas specific orbital and thermal energy, and $P$ the gas pressure\footnote{We point out that the expression given in \citet{Martire26} for $\mathcal{B}$ is missing a factor of 4 in the radiation energy density term. We have verified that including this factor does not quantitatively affect our results.}.
To elucidate the hydrodynamics of the unbound debris, we define as \emph{outflow} all the material with $v_{\rm r}>0$ and as \emph{wind} the material with $v_{\rm r}>0$ and $\mathcal{B}>0$ (i.e., unbound outflows). 
It is clear that material is expanding away from the black hole, escaping from the orbital plane and following increasingly radial trajectories over time, as can be seen by the streamlines. Although the outflow becomes more spherical during its evolution, anisotropies are evident at all times.

Fig.\ref{fig:Mdot_M} (left panel) shows the temporal evolution of the isotropic-equivalent mass loss rates for the outflow and the wind, respectively $\dot{M}_{\rm out}$ and $\dot{M}_{\rm w}$, computed as $\{ \theta, \varphi \}$- averages at $0.5a_{\rm mb}$ as described in Eq.\eqref{eq: Mdot iso}, where $a_{\rm mb}=r_{\rm t}^2/(2r_\star)$ is the semi-major axis of the most bound debris. The values are normalised by the Eddington accretion rate in maximally efficient conditions: $\dot{M}_{\rm Edd}\equiv L_{\rm Edd}/c^2$, where $L_{\rm Edd}\equiv 4\pi G M_{\rm BH}c/\kappa_{\rm p}$ is the Eddington luminosity computed with the average opacity at the photosphere at the peak time of the bolometric light curve $\kappa_{\rm p}$ as described in Sec.3.2 in \citet{Martire26}.
Overall, more material becomes unbound and the mass outflow/wind rates increase in time, remaining consistently super-Eddington. 
The stream side exhibits the highest mass outflow and wind rates, with dimensionless efficiencies\footnote{When computing $\zeta$, we use the actual mass loss rate through each sector, rather than its isotropic-equivalent value defined in Eq.~\eqref{eq: Mdot iso}. This ensures that $\zeta$ represents the fraction of the fallback rate that is actually carried away by the wind in that sector.}
$\zeta=|\dot{M}_{\rm w}/\dot{M}_{\rm fb}|$ reaching values $\zeta \approx 0.2$ by the end of the simulation (here $\dot{M}_{\rm fb}$ is the fallback rate of the bound debris under Keplerian motion, e.g. \citealt{Evans89}).  In other angular sectors, wind loss efficiencies $\zeta \ll 1$ at all times. 

After 2$t_{\rm fb}$, 67\% of the dynamically bound mass (i.e. $m_\star/2$) is in the outflow in the stream side, of which 2\% is unbound, with the majority in the middle stream side. In other words, the stream side features a bound, highly eccentric outflow next to the orbital plane, which is sheathed (in the middle stream) by an unbound wind. In contrast to these vigorous outflows, in the pericentre region (polar side), the outflow percentage decreases to 0.3\% (0.01\%), of which 38\% (96\%) is unbound.
Overall, only a small percentage (1\%) of the dynamically bound mass is in the wind, primarily concentrated in the middle stream side. This can be understood noticing that only 10\% of the stellar mass has returned to pericentre by the end of the simulation.

\subsection{Anisotropic wind}
The left column in Fig. \ref{fig:prof} shows $\{ \theta, \varphi\}$-averages (which are mass-weighted for the cases of gas density and radial velocity) radial profiles of the outflow and wind properties. Gas density profiles decline with radius as power laws with slope approximately $-2$ in all sectors, while advection luminosity profiles scale roughly as $r^{-2/3}$, consistent with the predicted behaviour of a radiation-driven wind \citep{Linial24, Martire26}. 
The density of the wind ranges between $10^{-7}$ and $10^{-11}$ g/cm$^{3}$, approximately $2-6$ orders of magnitude lower than the density of the core of the stream returning to apocentre, which is $\approx10^{-5}$ g/cm$^{3}$. Due to the higher densities of the stream side close to the orbital plane, its contribution to $\dot{M}$ exceeds that from the other regions by one order of magnitude. 
At small radii, the outflow across the pericentre side is dominated by the gas whose azimuthal velocity exceeds its radial velocity, indicating that the material resides only temporarily in this region before moving on to the stream side. This produces a spike in the radial profile of $\dot{M}$ (and $\rho$) on the pericentre side, which is absent from the other sectors. The poles and pericentre side are generally much more dilute than the stream side and reach higher velocities at large radii.

The typical wind velocity is between $10^{3}$ and $10^{4}$ km/s.  While this velocity range varies only moderately with angle, other wind properties such as $\rho$, $\dot{M}_{\rm w}$, and $L_{\rm adv}$ are highly anisotropic, as can be seen from the polar profiles in the right column of Fig. \ref{fig:prof}. 

This anisotropy may be driven by mass entrainment in the stream side. In this region, the wind deposits energy through dissipation, potentially unbinding initially bound but highly eccentric material slightly above the orbital plane. This interpretation is supported by the observed decrease in velocity profiles, which can be explained by newly unbound material decelerating previously ejected gas, similarly to what happens in mass-loaded outflows from accretion discs \citep{MetzgerStone16, Thomsen22}. Such behaviour is absent at higher latitudes on the stream side, where the velocity remains roughly constant, and in the pericentre and polar regions, where the debris expands more freely without significant interaction. 

Diamonds and dots in the panels of Fig.\ref{fig:prof} indicate, respectively, the position of the trapping radius $r_{\rm tr}$  (i.e  the radius at which the dynamical time $t_{\rm dyn}=r/v_{\rm r}$ equals the diffusive time $t_\text{diff}\equiv r \tau/c$) and that of the photosphere, computed in each region as medians among the lines of sight that contain an advective regime\footnote{We underline that we do not show wind profiles for the eccentric flow side since no wind material and no trapping radius are found there.}. 
The largest trapping radius is found in the middle stream side, resulting in a greater fraction of the energy being converted from radiation into kinetic energy (see, Sec.\ref{sec: lc}). 
\begin{figure}
    \centering
    \includegraphics[width=
    \linewidth]{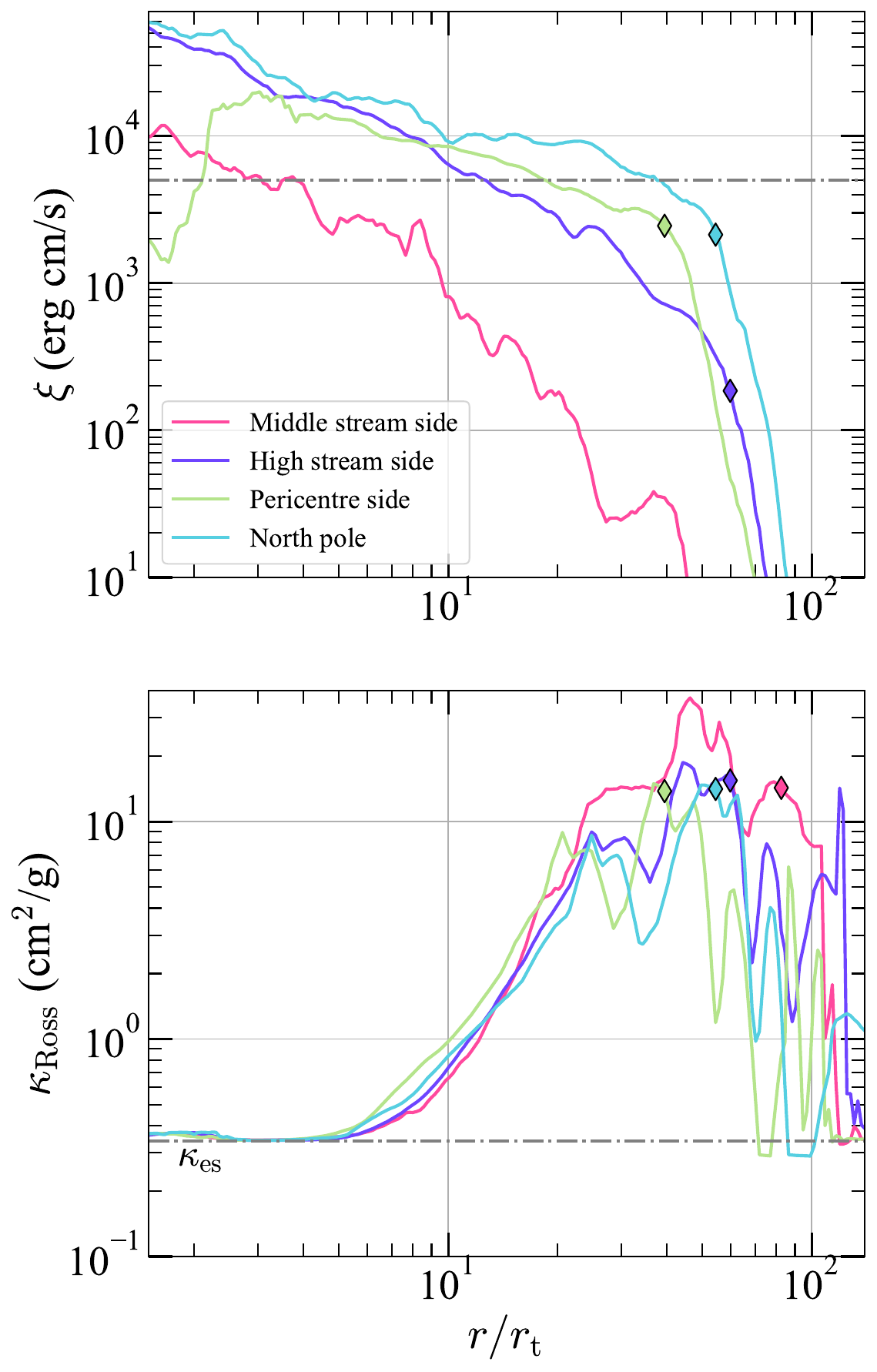}
    \caption{Wind radial profiles at the last available snapshot ($t=2.2t_{\rm fb}$), calculated as $\{ \theta,\varphi \}$- averages over spherical shells. 
    \emph{Upper panel}: Ionization parameter $\xi$. The horizontal dash-dotted line marks the threshold identified by \citet{Matt93, Matt96}, above which Thomson scattering is the dominant source of opacity.
    \emph{Lower panel}: Rosseland mean opacity. Diamonds mark the median $r_{\rm tr}$ across lines of sight within the selected regions that exhibit an advective wind. The horizontal dash-dotted line corresponds to the Thomson electron scattering opacity $\kappa_{\rm es}=0.34$cm$^2$/g. Unlike past investigations of SMBH TDEs \citep[e.g.][]{Roth16, SS24}, the IMBH TDE simulated here produces a wind dominated by absorption rather than scattering.}
    \label{fig:opac}
\end{figure}

\begin{figure*}
    \centering
    \includegraphics[width=\linewidth]{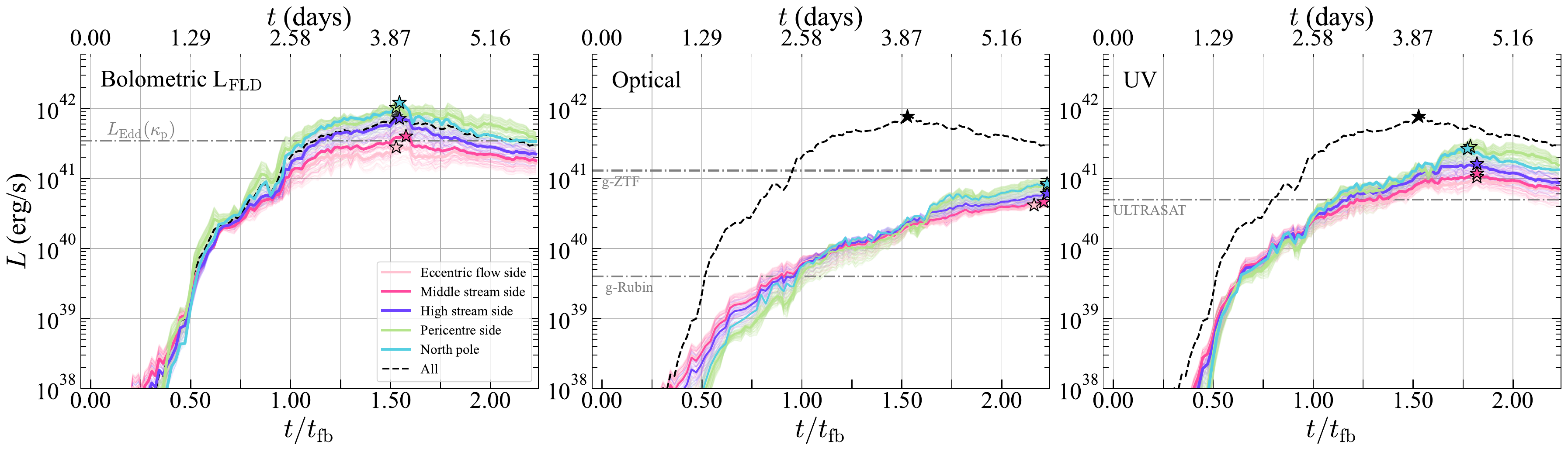}
        \caption{Bolometric luminosity (\emph{left panel}) and single-band light curves for optical  and UV (\emph{middle} and \emph{right panel}). Faint lines show the isotropic-equivalent luminosities emitted from each of the $N_{\rm obs}$ lines of sight (Eqs.~\eqref{eq:Lbol_weigh}, \eqref{eq:Lnu_weigh}), while thick coloured lines show their $\{\theta,\varphi\}$-averaged values within each sector. The $4\pi$-steradian bolometric light curve is shown as a dashed black curve in all the panels. 
        Stars indicate the peak of each averaged light curve. Horizontal dash-dotted lines show the minimum luminosities for detection in current and upcoming wide-field surveys, assuming a TDE at a distance of 100 Mpc. 
        Material in the polar and pericentre sides emits at 2-3 times the Eddington luminosity $L_{\rm Edd}$ (computed according to the average value of the Rosseland mean opacity $\kappa_{\rm p}\approx1.4$ cm$^2$/g at the time of the peak of the bolometric light curve, as explained in \citealt{Martire26}), shown as a horizontal dash-dotted gray line in the left panel; the stream side stays sub-Eddington. 
        Up to $1t_{\rm fb}$, the optical and UV emission is primarily driven by the stream side, after which the polar and pericentre sectors become the dominant source.  
        Overall, after $1.25t_{\rm fb}$ the UV and optical bands dominate the emission (see also Fig.\ref{fig: Xray lc}).}
    \label{fig: mono lc bands}
\end{figure*}

\subsection{Absorption dominated  wind}
As found in \citet{Roth16}, the wind is initially optically thick and scattering-dominated. However, we find that, after the peak of the light curve,  absorption becomes increasingly important, first on the stream side, then in the polar regions, and finally on the pericentre side. By the end of the simulation, the wind is generally absorption dominated (see lower panel in Fig.\ref{fig:opac}). This is likely a consequence of the higher densities compared with outflows in TDEs by $10^6M_\odot$ BHs,  together with the complex dependence of opacity on density and temperature, which precludes a description of absorption by Kramers’ law alone (see footnote 4 in \citealt{MetzgerStone16}).
To assess our findings, we compute the ionization parameter $\xi$ introduced by \citet{Miller15xi} as  $\xi \equiv 4\pi F/n_{\rm e}$ (see \citet{Matt93, Matt96} for an early application in accretion discs),
where $F$ is the ionizing energy flux, $n_{\rm e}\approx\rho/\mu$ the number density of electrons in a spherical wind, $\mu\approx1.3m_{\rm p}$ the mean particle mass for solar composition, and $m_{\rm p}$ the mass of a proton. This parameter quantifies the number of ionizing photons per electron. \citet{Matt93, Matt96} showed that when $\xi\gtrsim 5000$ erg cm/s, the gas is strongly ionized,  which means that Thomson scattering, rather than absorption, dominates the opacity. We evaluate $\xi$ 
approximating the ionizing energy flux with the Planck function at the mean radiation temperature of each sector, integrated in frequency from 13.6~eV to infinity.
We find values $\leq3000$ erg cm/s (see upper panel in Fig.\ref{fig:opac}) in regions where the absorption opacity is high.
While densities differ by at least one order of magnitude between the
 middle stream side and the other regions, temperatures differ only by a factor of a few ($T_{\rm rad}(r_{\rm tr})\approx10^4$~K for the middle stream, compared with $T_{\rm rad}(r_{\rm tr}) \approx4-5\times10^4$~K for the others). This reduces the number of ionizing photons, and thus $\xi$, in the  middle stream relative to the other regions, but opacities remain similar across all regions due to their stronger dependence on temperature rather than on density.

\begin{figure}
    \centering
    \includegraphics[width=\linewidth]{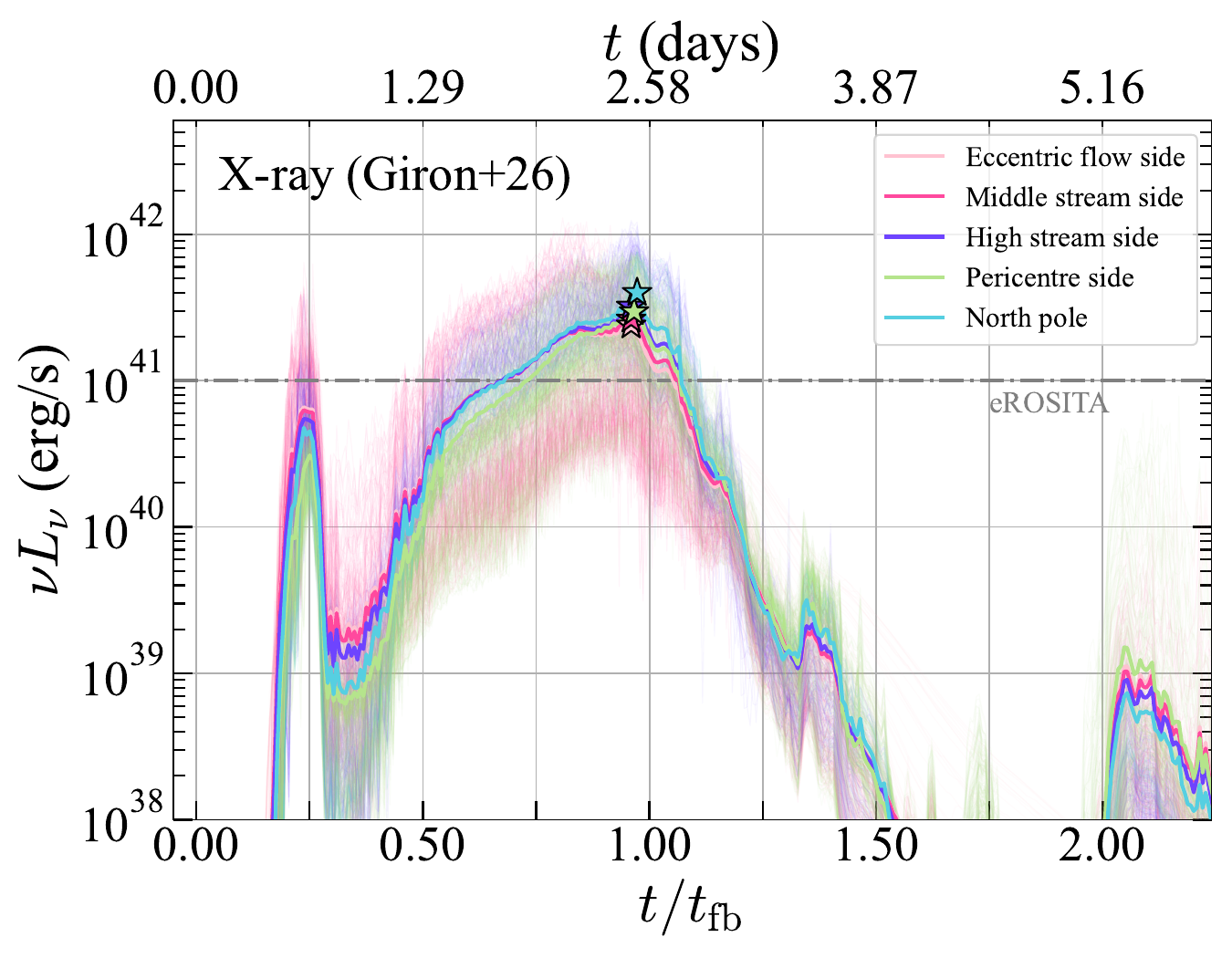}
        \caption{X-ray light curve from the multigroup simulation presented in \citet{Giron26}. Faint lines show the isotropic-equivalent luminosities emitted from each of the $N_{\rm obs}$ lines of sight (Eq.~\eqref{eq:Lnu_weigh}), while thick coloured lines show their $\{\theta,\varphi\}$-averaged values within each sector.  Stars indicate the peak of each average light curve. 
        An X-ray flare is emitted around $1t_{\rm fb}$, above the \emph{eROSITA} detection threshold at 100~Mpc (horizontal dash-dotted line).} 
    \label{fig: Xray lc}
\end{figure}

\begin{table*}
\centering
\caption{Luminosity ratios at three times for the five angular regions.}
\label{tab:ratios}
\small
\begin{tabular}{lcccccccccc}
\toprule
Time ($t_{\rm fb}$)
& \multicolumn{2}{c}{Eccentric flow side}
& \multicolumn{2}{c}{Middle stream side}
& \multicolumn{2}{c}{High stream side}
& \multicolumn{2}{c}{Pericentre side}
& \multicolumn{2}{c}{North pole}\\
\cmidrule(lr){2-3}\cmidrule(lr){4-5}\cmidrule(lr){6-7}\cmidrule(lr){8-9}\cmidrule(lr){10-11}
& L$_{\rm X}$/L$_{\rm op}$ & L$_{\rm op}$/L$_{\rm UV}$
& L$_{\rm X}$/L$_{\rm op}$ & L$_{\rm op}$/L$_{\rm UV}$
& L$_{\rm X}$/L$_{\rm op}$ & L$_{\rm op}$/L$_{\rm UV}$
& L$_{\rm X}$/L$_{\rm op}$ & L$_{\rm op}$/L$_{\rm UV}$
& L$_{\rm X}$/L$_{\rm op}$ & L$_{\rm op}$/L$_{\rm UV}$ \\
\midrule
1.0 & 22.2 & 0.24 & 26.2 & 0.24 & 35.2 & 0.21 & 45.6 & 0.15 & 45 & 0.18 \\
1.5 & $5\times10^{-3}$ & 0.27 & $5\times10^{-3}$ & 0.24 & $4.0\times10^{-3}$ & 0.21 & $3.1\times10^{-3}$ & 0.16 & $2.9\times10^{-3}$ & 0.18 \\
2.2 & $8\times10^{-3}$ & 0.61 & $6\times10^{-3}$ & 0.64 & $3.9\times10^{-4}$ & 0.70 & $3.5\times10^{-3}$ & 0.52 & $1.8\times10^{-3}$ & 0.63 \\
\bottomrule
\end{tabular}
\end{table*}

\subsection{Light curves}
\label{sec: lc}

Figure~\ref{fig: mono lc bands} shows the bolometric (\emph{left panel}), optical (\emph{middle panel}; $1.6767$-$3.358$ eV), and UV (\emph{right panel}; $3.358$-$7.7488$ eV) light curves, while the X-ray emission ($300$-$2\times10^4$ eV) is presented in Fig.~\ref{fig: Xray lc}. In all panels, faint lines show the emission seen by individual observers, while thick lines represent the $\{\theta,\varphi\}$-averaged light curves among observers within each region.
For the bolometric luminosity, $L_{\rm FLD}$, the emission seen by each observer $i$ accounts for contributions from the other lines of sight through the cosine-weighted average
\begin{equation}
\label{eq:Lbol_weigh}
    \frac{\sum_{k=0}^{N_{\rm obs}}\max(0,\cos\delta_{k,i}) L_{\rm FLD,k}}{\sum_{k=0}^{N_{\rm obs}}\max(0,\cos\delta_{k,i})}.
\end{equation}
where $\delta_{k,i}$ is the angular separation between the $k$-th line of sight and observer $i$. The single-band luminosities are computed analogously, following Eq.~\eqref{eq:Lnu_weigh}. For the X-ray, we adopt the results from \citet{Giron26}, where the same simulation setup is used but with (i) lower resolution, and (ii) a multigroup diffusion approximation rather than a gray one. This choice is motivated by the limitations of the gray diffusion approach, which relies on Rosseland mean opacities, and whose weighting function favour lower frequencies. As a result, X-ray opacities are underestimated, leading to artificially enhanced luminosities in that band\footnote{We emphasize that the optical and UV bands are in good agreement between gray and multigroup approximation. This can be seen in Appendix \ref{app:MG}, which also quantitatively demonstrates the poorer performance of grey radiation transfer at estimating X-ray emission for $t \gtrsim 1 t_{\rm fb}$.}.

The lower bolometric luminosity of the middle stream side is consistent with its higher mass-loss rates and larger trapping radii, which are associated with a more extended advective zone. As a result, a larger fraction of its internal, radiation-dominated energy is converted into kinetic energy rather than escaping as radiation to infinity, resulting in sub-Eddington emission. By contrast, the polar and pericentre sides reach peak luminosities of 2 to 3 times the Eddington limit.  
At early times, the X-ray band is a large fraction of the total bolometric luminosity, with the polar regions reaching the highest X-ray luminosities\footnote{The first X-ray peak at $t\approx0.25 t_{\rm fb}$ is likely an artifact of spurious numerical dissipation and is therefore excluded from our interpretation.}.  As the mass-loss rates increase (Fig.\ref{fig:Mdot_M}), the wind becomes more opaque, and reprocesses the radiation produced in the inner regions to lower frequencies \citep{Giron26}. Before $t \approx 1.25t_{\rm fb}$, X-ray luminosities exceed optical/UV luminosities by a factor of $\approx 10$ or more, but after this time, the UV and optical bands become dominant in all sectors.  
As the optical emission increases, these ratios decrease substantially to $\sim 10^{-3}-10^{-2}$ (see Table~\ref{tab:ratios}), reaching their lowest values at late times in and around the polar region. In contrast, the optical-to-UV ratio increases with time, although the UV emission remains a factor of a few stronger than the optical emission throughout.
The shape and time of the peak of the bolometric light curve (dashed black line) resemble that of the UV one, while the optical light curves peak $\approx 0.5t_{\rm fb}$ later.
The single-band light curves show anisotropy patterns that are qualitatively similar to the bolometric anisotropy, though less pronounced. Up to $1t_{\rm fb}$, the UV and optical emission is primarily driven by the stream side, after which the polar and pericentre sector becomes the dominant source. 
We note that, prior to the angular weighting, the highest luminosities are observed from the polar and pericentre directions, with the stream side being systematically fainter. The angular weighting partially smooths out these differences: the bright polar emission enhances the stream side luminosity, while contributions from the fainter stream side reduce the luminosity observed from the poles. The resulting observed emission therefore retains a dependence on viewing angle, but with a reduced contrast among the different sectors.

\begin{figure*}
\centering
\includegraphics[width=\linewidth]{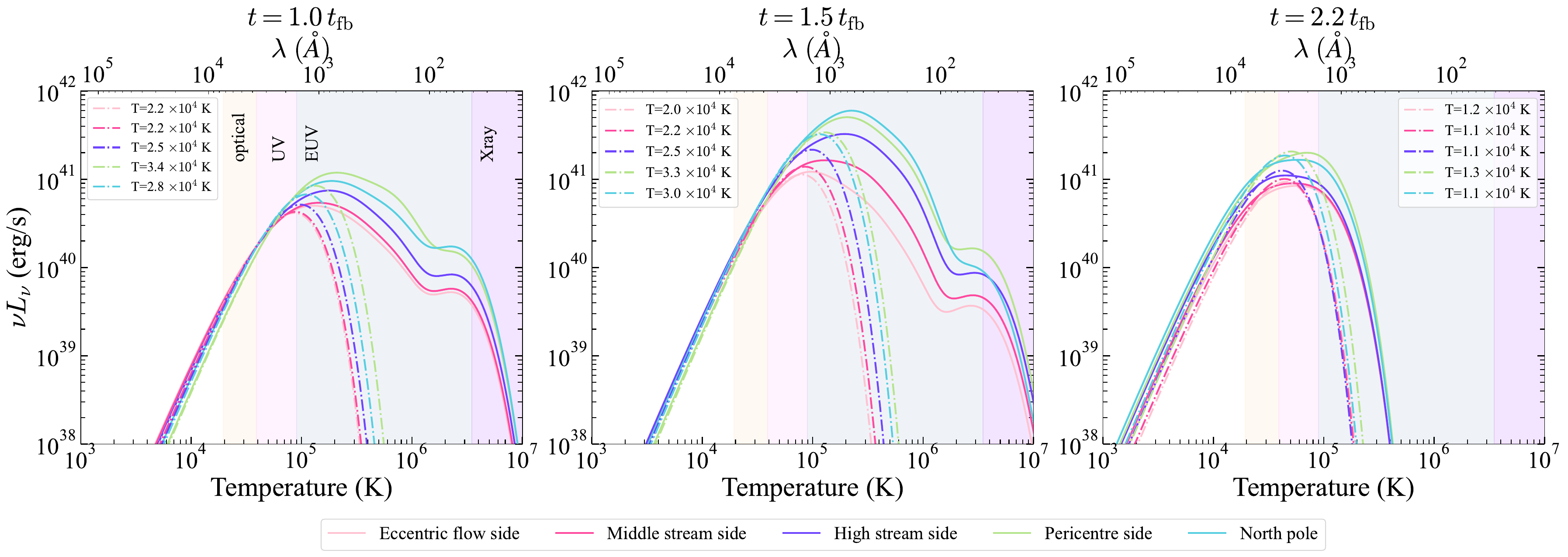}
\caption{SED at $t=1 t_{\rm fb}$ (\emph{first panel}), at the time of the peak of the bolometric light curve $ 1.5 t_{\rm fb}$ (\emph{second panel}), and at the last time available $ t=2.2 t_{\rm fb}$ (\emph{third panel}) computed in post-processing as in Eq.\eqref{eq:Lnu_weigh}, with a $\{ \theta,\varphi \}$-average over each sector (labelled by solid coloured lines; see lower legend). The upper x-axis shows the corresponding wavelengths to the temperature on the lower x-axis, computed as $\lambda=ch/(k_{\rm B}T)$.
The dash-dotted lines represent the black-body fits using the optical-UV bands from ZTF (r, g, i) and \emph{Swift}/UVOT (v, b, u, uvw1, uvm2, uvw2). The fitted temperatures (see in-panel legends) remain almost constant in time before the peak of the bolometric light curve, and then decreases by up to a factor of 2.
Emission shifts from X-ray/EUV to UV/optical, with the majority of light coming from the polar and the pericentre sides.}
\label{fig:spectra}
\end{figure*}
\begin{figure*}
    \centering
    \includegraphics[width=\linewidth]{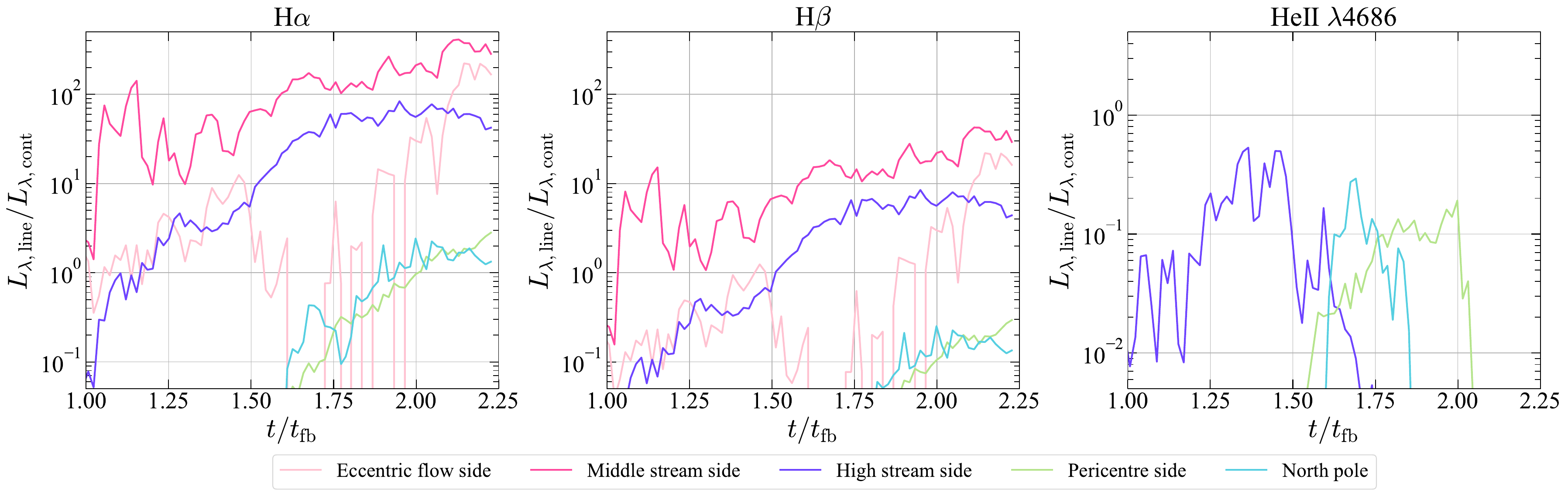}
    \caption{Evolution of line-to-continuum luminosity ratios for H$\alpha$ (\emph{left panel}), H$\beta$ (\emph{middle panel}), and HeII ~$\lambda4686$ (\emph{right panel}), evaluated in post-processing at the photosphere. While He emissivities are lower than the continuum and thus challenging to observe, H$\alpha$ and H$\beta$ lines are detectable for observers on the \emph{stream side}, suggesting that future optical spectroscopy of IMBH TDEs may categorize them as ``TDE-H'' flares \citep{Arcavi14, vanVelzen21, Yao23}.}
    \label{fig:lines}
\end{figure*}
\subsection{SEDs}
We use Eq.~\eqref{eq:Lnu_weigh} to compute the SED at three representative times, shown in Fig.~\ref{fig:spectra} as $\{ \theta,\varphi \}$-average for each sector. The resulting spectra are a superposition of multiple blackbody components, reflecting the anisotropic structure of the outflow. The dash-dotted curves indicate single–blackbody fits obtained using only the optical and UV bands from ZTF (r, g, i; see \citealt{ztf}) and \emph{Swift/UVOT} (v, b, u, uvw1, uvm2, uvw2; see \citealt{Swift}). As found in previous studies \citep{Roth16, Thomsen22, Parkinson22, Mockler26}, these fits systematically underestimate the bolometric luminosity by at least an order of magnitude at early times, because the SED peaks in the 
unobservable EUV.  At later times, the emission shifts from X-ray/EUV to UV and optical, and the spectra are better fit by single-temperature blackbodies by the end of the simulation (though such a fit still underestimates the true bolometric luminosity).
The effective temperature from the optical/UV fit remains nearly constant before the peak of the bolometric light curve, and then decreases by up to a factor of 2, reaching $T_{\rm eff}\approx 10^4$K, below observed UV/optical temperature fits in SMBH TDEs \citep{Yao23}.
Most of the emission is observed from the pericentre and polar sides, with the pericentre exhibiting the highest luminosities also because of the angularly weighted contributions from the other lines of sight, as noted in Sec.\ref{sec: lc}. 

\begin{figure*}
    \centering
    \includegraphics[width=\linewidth]{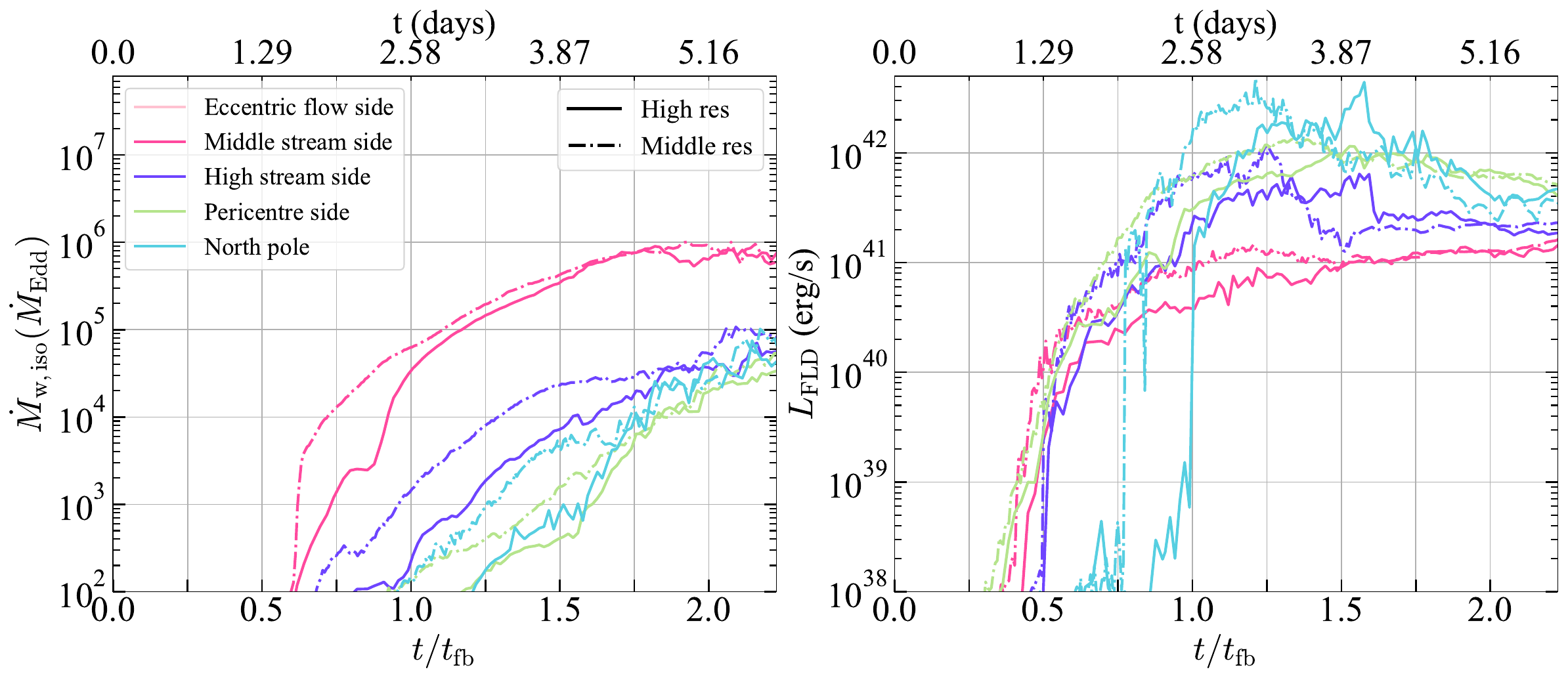}
    \caption{Resolution tests for time evolution of (isotropic-equivalent) wind mass-loss rate $\dot{M}_{\rm w, iso}$ (solid lines in left panel in Fig.\ref{fig:Mdot_M}) and bolometric luminosity $L_{\rm FLD}$ (left panel in Fig. \ref{fig: mono lc bands}). Solid and dash-dotted lines represent the curves obtained with the resolution used in this work (\emph{High res}) and with a lower resolution (\emph{Middle res}) taken from \citet{Martire26}, respectively. Resolutions agree within a median factor over time of $\approx 1.5$ in every sector from $1.5t_{\rm fb}$ onward.}
    \label{fig:ML conv}
\end{figure*}

\subsection{Line luminosities}
We compute specific line luminosities (Eq. \ref{eq:Lline}) for H$\alpha$, H$\beta$, and HeII~$\lambda$4686 transitions.
We treat H, and HeII as hydrogenic (with $\chi=13.6$eV and $\chi=54.4$eV, respectively) with $g_ n=2n^2$, $\chi_n = \chi (1 - n^{-2})$, and we adopt a cut-off $n_{\max}=30$. The partition function of neutral helium is computed using the first five entries from the NIST dataset\footnote{\url{https://physics.nist.gov/PhysRefData/Handbook/Tables/heliumtable5.htm}.} and $\chi=24.6$~eV.
We approximate $\Delta r\approx r_{\rm ph}$ \citep{Aspegren26} and define $\Delta\lambda=\lambda_0 v(r_{\rm ph})/c$, with $\lambda_0$ being the line centre wavelength.
Atomic data ($\lambda_0$ and Einstein $A$ coefficients) are taken from standard tabulations (see Appendix \ref{app: lines}). 
Figure~\ref{fig:lines} shows the $\{ \theta, \varphi\}$-median ratios of the specific line luminosities to the continuum luminosity (Eq.~\ref{eq:Lcont}) in the different angular regions.

The hydrogen lines exhibit a pronounced viewing-angle dependence. In particular, both H$\alpha$ and H$\beta$ are strongest in the densest regions (i.e., the stream side), whereas their luminosities remain much weaker toward the poles and the pericentre side. This anisotropy becomes particularly pronounced at late times, when the Balmer-line luminosity can exceed the continuum luminosity by more than an order of magnitude along the stream, while remaining comparable to or below the continuum in other directions.
We note that, in agreement with observations \citep{Charalampopoulos22}, the hydrogen lines exhibit a time lag with respect to the continuum luminosity, continuing to increase after the peak of the light curve. Detectable H$\alpha$ emission with  temporal variability has also been reported in some optically selected TDE candidates \citep{Arcavi14, Holoien16b}.
By contrast, the helium lines are most prominent around the peak of the light curve and preferentially at or near the poles, where the radiation field is harder. Nevertheless, their luminosities remain below the continuum luminosity throughout the evolution. At late times, as the photosphere expands, the helium emission becomes progressively weaker, reaching $\approx 0.1 L_{\rm cont}$. The progressive weakening of the helium lines is qualitatively consistent with the stratified-atmosphere proposed by \citet{Roth16}, who found that the helium photosphere lies deeper than the hydrogen one. Our results suggest that, as the photosphere expands, this stratification may contribute to the increasingly weak helium emission relative to the continuum.

\subsection{Resolution tests}
The compression of returning streams at the pericentric nozzle is likely to be locally under-resolved: fully resolving the peak compression and shocks in this region would require quite high vertical resolution, which is more easily achieved in lower dimensionalities.  For example, the resolved 1D simulations of \citet{Andalman26} find that, at maximum compression, the stream collapses vertically to heights $\sim 10^{-4}R_\odot$.  The resolved 2D simulations of \citet{Bonnerot22} find that maximum compression is attained at a height $\sim 10^{-3} R_\odot$.  In contrast, \citet{SS24} map pre-return streams from their global 3D simulation (which is under-resolved at pericentre) into a resolved 2D simulation (following the \citealt{Bonnerot22} framework) and find that compression is reduced by prior (resolved) 3D effects which pre-heated the stream, so that nozzle compression reaches its maximum extent at $\sim 10^{-2} R_\odot$.  In this work, we find a maximum compression of $\sim 10^{-1}R_\odot$. Although we caution that these are not straightforward comparisons, as the parameters (e.g. $M_{\rm BH}, r_{\rm p}$) and microphysics (e.g. recombination effects, radiation transport) of the quoted simulations frequently differ from each other, this progression is reminiscent of older debates on the ``pancaking'' of a star during the first pericenter passage, in which tidal disruption occurs.  1D analytical theory \citep{Carter83} and hydrodynamical simulations \citep{Brassart08} both found extreme degrees of compression which were not borne out in (resolved) 3D simulations, where multidimensional effects reduced compression (relative to the 1D case) by orders of magnitude \citep{Guillochon09, Coughlin22}.  More recent 3D simulations have used a GPU-accelerated smoothed-particle-hydrodynamics code to resolve an extreme degree of compression (down to $\sim 10^{-4} R_\odot$) in a nozzle shock \citep{Kubli26}, although the isentropic equation of state used here may overestimate the pericentric compression due to the absence of shock heating.

To assess whether vertical under-resolution at pericentre affects our conclusions regarding the nozzle shock, we perform the diagnostic proposed by \citet{Hu26}. At the same evolutionary stage and within a comparable spatial region\footnote{i.e., $ r \leq2200 r_{\rm g},~\varphi\in[\varphi_{\rm md}-4.5^\circ, \varphi_{\rm md}+4.5^\circ]$ and $t\approx0.3t_{\rm fb}$, with $\varphi_{\rm md}$ being the azimuthal angle of the point of maximum dissipation.}, we find a ratio of internal to kinetic energy of $\approx 8\times10^{-5}$, below their analytical estimate for the fractional energy dissipation during nozzle passage, $(m_\star/M_{\rm BH})^{2/3}\approx10^{-3}$. Measuring the dissipated energy directly from the simulation, we obtain a fractional dissipation of $\approx6\times10^{-3}$.
Although this exceeds the analytical estimate, the latter should be regarded as a lower limit and can be substantially enhanced, for example, by recombination puffing up the debris stream.
The discrepancy is therefore not necessarily indicative of excessive numerical dissipation, although some contribution from the numerical treatment of the nozzle shock cannot be ruled out. In either case, the dissipated energy remains small compared to the stream's kinetic-energy budget. Therefore, while the detailed thermodynamics of the nozzle may not be numerically converged, the possible overestimate of the dissipated energy is not sufficiently large to alter the subsequent global evolution.

\begin{figure}
    \centering
    \includegraphics[width=0.8\linewidth]{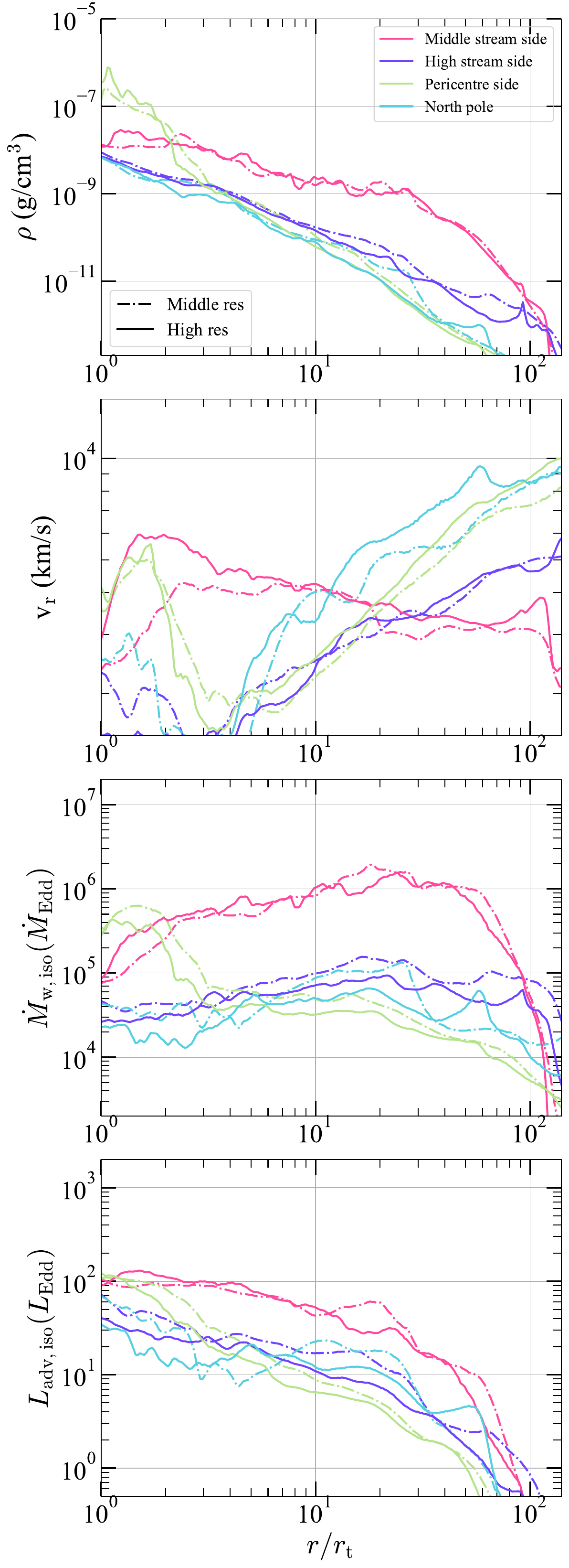}
    \caption{Resolution test for wind radial profiles shown in Fig.\ref{fig:prof}. Colours and linestyles scheme are the same of Fig.\ref{fig:ML conv}. The two resolutions show the same qualitatively behaviour and converge within a median factor of $\approx 2$.}
    \label{fig:prof convT}
\end{figure}

\citet{Martire26} already showed that the formation and behaviour of the outflow are insensitive to numerical resolution, with the primary difference being a temporal delay in the onset of the wind in the higher resolution run. To further assess the robustness of our analysis, we compare the results presented in Sec.~\ref{sec: res} - hereafter referred to as \emph{High res}- with those obtained using the same methods applied to a simulation with identical initial parameters but with the initial number of cells reduced by a factor of four relative to the setup adopted in this work -  hereafter referred to as \emph{Middle res}, following the nomenclature introduced by \citet{Martire26} for consistency.
The temporal delay mentioned above is evident in Fig.\ref{fig:ML conv}, in particular for the pole and pericentre side. Nevertheless, mass loss rates and luminosities converge toward similar values at late times, with the different resolutions agreeing within a median factor over time of $\approx 1.5$ in every sector from $1.5t_{\rm fb}$ onward. Comparing the radial profiles at the last available time ($t = 2.2\,t_{\rm fb}$), we note that they agree within a median factor of $\approx 2$ (see Fig.~\ref{fig:prof convT}). Overall, the qualitative behaviour is independent of resolution. The pericentre and pole regions are similar to each other, with low densities and mass loss rates and high emitted luminosities. The stream side is generally denser and slower with more evident anisotropies: $\dot{M}_{\rm w}$ increases and the bolometric luminosity decreases with increasing latitude (i.e. towards the orbital plane).

\section{Discussion}
\label{sec: disc}
The origin of optical and UV emission in TDEs remains an open question. Several theoretical frameworks have been proposed, 
with substantial overlap between their underlying physical mechanisms and no uniform terminology across the literature.
The most widely discussed mechanisms for the early-time emission are (i) dissipation of orbital energy in shocks, whose radiation can be locally thermalised or further reprocessed at larger scales to longer wavelengths if they are optically thick \citep{Piran15, Jiang16, Bonnerot17, Ryu20, SS24, Martire26}, and (ii) reprocessing of hard radiation from a central accretion flow to softer frequencies in an extended, optically thick medium, which may be quasi-static \citep[e.g.][]{Loeb97, Roth16, Metzger22}, or may be outflowing \citep[e.g.][]{MetzgerStone16, Dai18, Price24, Huang25}. We underline that reprocessing in an outflow can coexist with shocks, since the outflow itself may arise from different mechanisms. In particular, the outflow can be driven either by a super-Eddington accretion wind powered by an inner accretion disc \citep{Strubbe09, LodatoRossi11}, or alternatively by a super-Eddington energy dissipation rate associated with shocks \citep{Jiang16}.
In this work, we have investigated the evolution of the wind self-consistently produced by the nozzle shock following the disruption of a $0.5 M_\odot$ star by a $10^4 M_\odot$ IMBH.

\subsection{Comparison with previous simulations}
\label{sec:disc sim}
Across existing simulations, whether the outflowing material remains as a bound envelope or becomes unbound depends primarily on the interplay between the initial binding energy of the debris and the energy input from shocks and radiation. 
In simulations producing extended bound structures, the energy imparted to most of the debris remains insufficient to overcome its binding energy. Physical and numerical conditions vary considerably in these calculations, but include reduced black-hole-to-star mass ratios \citep{Shiokawa15}, initially bound stellar orbits \citep{Hayasaki16, Bonnerot16}, or inefficient dissipation of orbital energy through shocks \citep{Price24}.
In contrast, a number of simulations found substantial unbound components, generally when weakly bound debris undergoes strong energy injection. 
This energy can be supplied by pericentre compression \citep{Ayal00}, self-crossing shocks associated with deep, relativistic encounters \citep{Sadow16}, or radiative heating at sufficiently high mass-flow rates and near-Eddington luminosities (\citealt{Jiang16}, \citealt{Bonnerot21}, case A, \citealt{Huang24}). 
Our simulation falls in this latter regime, as the super-Eddington dissipation at the nozzle shock provides sufficient energy to unbind an increasing fraction of the initially weakly bound debris. Higher mass loss rates correspond to enhanced wind densities, leading to increased reprocessing of high-energy radiation and a higher optical-to-X-ray flux ratio at later time \citep{Giron26}. 

Despite differences in the dynamical origin and binding energy of the reprocessing material, its radiative properties share several features with previous models.
Consistent with \citet{Mockler26}, higher-energy bands peak earlier in time, while lower-energy emission peaks progressively later. Most of the reprocessed luminosity emerges in the UV, and the optical light curves peak with a delay of at least $0.5t_{\rm fb}$ with respect to the bolometric light curve. 
The angular dependence of the emission also agrees with previous models \citep{Dai18, Thomsen22} in presenting  notable variations in wind density, velocity, and mass-loss rate with inclination. In addition, we identify a strong dependence on the azimuthal direction, with the the stream side being slower and denser while the pericentre side is similar to the poles.
The poles and the pericentre side results brighter, while the luminosity is sub-Eddington when observed from the stream side, especially close to the midplane.
However, unlike \citet{Dai18, Thomsen22}, we do not find an increase in the optical-to-X-ray flux ratio toward the orbital plane. This discrepancy may arise from significant contamination from material outside the considered line of sight \citep{Parkinson25} as well as the anisotropic, shock-driven nature of the wind. In our case, the strong azimuthal dependence of the wind implies that approaching the orbital plane does not necessarily correspond to encountering a reprocessing layer of increasingly larger optical depth. This azimuthal asymmetry is a natural consequence of the localized origin of the wind at the nozzle shock. In the black hole frame, the velocity imparted by the nozzle shock adds to the orbital velocity of the returning debris. The outflow therefore retains a preferred direction associated with the orbital motion at the nozzle, unlike an axisymmetric disc wind \citep{Dai18, Thomsen22}, where material is launched from all azimuths with no preferred azimuthal direction.

\subsection{Observational implications}
IMBH TDEs would offer a unique probe of these mid-mass compact objects, enabling tests of astrophysical scaling relations such as the $M-\sigma$ relation \citep{Msigma} at the poorly understood \citep{Greene20} low mass end. Although IMBH TDE detection is challenging -- because their emission is fainter and often below the sensitivity limits of current telescopes -- several candidates have been reported previously.  Interestingly, while optically selected TDE samples include IMBHs down to $M_{\rm BH} \sim {\rm few} \times 10^5 M_\odot$ \citep{Angus22, Yao23}, only X-ray selected TDE samples \citep{Maksym14, Lin18, Sazonov21, Wen21, Cao23} contain candidate flares from $M_{\rm BH} \sim 10^4 M_\odot$ IMBHs.  We now assess whether this dearth of smaller IMBHs in optically selected TDE samples is understandable.

To compare the detectability by survey capabilities, we estimate the detection thresholds of current and upcoming wide-field surveys at a fiducial distance of $d=100$Mpc and show them as dash-dotted horizontal lines alongside our single-band light curves in Figs.\ref{fig: mono lc bands} and \ref{fig: Xray lc}.
We compute the limiting luminosity as $L=4\pi d^2F$, where $F$ is the limiting flux in a given band. In the X-ray regime, the minimum detection thresholds are $\approx10^{41}$ erg/s for \emph{SRG/eROSITA} (considering exposure times of order 100~s and adopting the variability selection criteria of \citet{Grotova25a, Grotova25}) and $\approx 3.5 \times 10^{43}$ erg/s for {\it Einstein Probe/WXT} (based on Fig.6 of \citealt{TubinArenas24} and Table 1 of \citealt{Yuan25}, and considering exposure times of 1 ks). In the optical g-band, adopting a limiting magnitude of 20.8 for ZTF (Fig.6 in \citealt{ztf}) and 24.8 for the Vera Rubin Observatory \citep{Bianco22},  we obtain minimum detectable luminosities of $\approx1.3 \times 10^{41}$ erg/s and $4 \times 10^{39}$ erg/s, respectively, assuming $\lambda= 480$ nm. 
In the UV, a limiting magnitude of 22.4 at 260 nm for \emph{ULTRASAT} \citep{Shvartzvald24} corresponds to a threshold of $\approx5 \times 10^{40}$ erg/s.  
We can more precisely estimate the maximum detection horizon $d_{\rm h}$ for each survey for a TDE similar to the one simulated here. We assume $H_0=70$km~s$^{-1}$~Mpc$^{-1}$, $\Omega_{\rm M}=0.3, ~\Omega_\Lambda=0.7$, and determine $d_{\rm h}$ from the peak in-band luminosity of each sector (marked by stars in Figs.~\ref{fig: mono lc bands} and \ref{fig: Xray lc}). For the optical and UV bands, we compare the resulting observed flux density, $F_{\nu,\rm obs}$, with the survey limiting AB magnitude, using $m_{\rm AB}=-2.5\log_{10}[F_\nu/(3631,{\rm Jy})]$. We find that Rubin can detect the early-time emission out to $\approx300$ Mpc when viewed from the eccentric-flow side and $\approx790$ Mpc from the North pole, while the corresponding horizons for \emph{ULTRASAT} are $\approx90$ and $\approx340$ Mpc, respectively. In X-rays, adopting a TDE detection threshold of $F_{\rm obs}\approx1\times10^{-13}$ erg~cm$^{-2}$~s$^{-1}$ in the $0.2-2.3$ keV band \citep{Grotova25a,Grotova25}, we obtain an \emph{eROSITA} horizon of $\approx140-180$ Mpc.

For ZTF, the corresponding detection horizons are $\approx 50$ and $\approx 130$ Mpc for the eccentric-flow and north pole sides, respectively.  These modest horizon distances are compatible with the non-discovery of IMBH TDEs (from $M_{\rm BH} \sim 10^4 M_\odot$) in the ZTF sample, provided IMBH TDE rates\footnote{While the rate of TDEs from IMBHs is uncertain, and some past work has indicated that IMBHs may disrupt stars at a {\it higher} rate than SMBHs \citep{Stone16}, recent loss cone modelling suggests that TDE rates peak near $M_{\rm BH} \sim 10^6 M_\odot$ and decline steeply for IMBHs \citep{Chang25, Hannah25}.} are not much greater than SMBH TDE rates (for comparison, the closest detected SMBH TDE in the ZTF sample, AT2019qiz, is located at $d \approx 65$ Mpc; \citealt{Yao23}). 
While our estimates suggest that the ZTF survey was marginally capable of finding TDEs from low-mass IMBHs, 
Rubin and \emph{ULTRASAT} are expected to substantially extend the volume over which the early optical/UV emission from IMBH TDEs can be detected. In X-rays, \emph{eROSITA} is able in principle to detect the early time flare that we predict, while \emph{Einstein Probe} would only be able to do so for unrealistically nearby IMBH TDEs.

So far we have only discussed flux limits on IMBH TDE detection; however, the more rapid temporal evolution of IMBH TDEs suggests that detection may also be strongly cadence-limited for many surveys.  \emph{eROSITA}, with its typical $\sim 6$-month cadence, is the most notable example of this; as the early IMBH emission may evolve on timescales much shorter than the interval between successive observations, a substantial fraction of otherwise detectable events could be missed.  This issue may be more salient in the optical/UV than in the X-ray bands, as X-rays are (after the first mass return) sourced by inner disk emission \citep{Cannizzo90, vanVelzen19}, which can evolve quite slowly with time \citep{MummeryBalbus20, Alush25}.  In contrast, IMBH wind reprocessing of the type seen in our simulations is inherently time-limited: as the mass fallback rate drops, the wind mass-loss rate will also decrease, and eventually the effective optical depth will become too low to thermalize optical photons \citep{Matsumoto21}.

At the end of our simulation, $t=2.2t_{\rm fb}$, the wind emerging in the middle stream side contains the bulk of the unbound mass ($\approx2\times10^{−3}M_\odot$) with a characteristic velocity $v\approx0.012c$ beyond apocentre and kinetic energy $E_{\rm kin}\approx 4\times10^{47}$~erg. The polar wind is faster ($v\approx0.03c$), but contains only $\approx3\times10^{−5}M_\odot$, with $E_{\rm kin}\approx 10^{46}$~erg. Since both the unbound mass and kinetic-energy flux are still evolving when the simulation terminates, these values should be regarded as lower limits to the final ejecta mass and energy.
Once the wind is launched, it can propagate to larger radii and interact with the circumnuclear medium (CNM), driving shocks capable of producing synchrotron radio emission \citep{LuBonnerot20}. 
The properties of our simulated outflow are comparable to those inferred for several radio-emitting thermal TDEs. A relevant case is AT2020vwl, for which radio observations revealed a non-relativistic outflow launched around the time of the optical flare. Assuming a spherical geometry (and the energy carried
by the electrons being 0.1 times the total energy), \citet{Goodwin23} inferred an outflow mass $M_{\rm w}\approx10^{−3}M_\odot$, kinetic energy $E_{\rm kin}\approx 10^{48}~{\rm erg}$ and velocity $v\approx0.03c$. They further argued that the outflow was likely produced by material ejected during stream-stream collisions, rather than requiring an accretion-powered wind, making it particularly relevant comparison to our nozzle-shock-driven wind.
ASASSN-14li provides another useful comparison. Its radio emission was interpreted as arising from a non-relativistic outflow with $M_{\rm w}\approx3\times10^{−5}-7\times10^{−4}M_\odot, E_{\rm kin}\approx (4-10)\times10^{47}~{\rm erg}, v\approx0.04-0.12c$ \citep{Alexander16}. The mass scale is therefore comparable to that of our simulated wind, particularly between the polar and pericentre components, although the ASASSN-14li outflow is faster and it has been associated to a (super-Eddington) accretion-driven wind around a $10^6M_\odot$ SMBH.
At later times, radio observations reveal a broader population of non-relativistic TDE outflows, with inferred characteristic velocities of $0.02-0.15c$ and kinetic energies of $10^{47}-10^{49}$~erg \citep{Cendes24}. 
These comparisons suggest that shock-driven outflows in IMBH TDEs as the one simulated here can provide the mass, velocity, and kinetic-energy reservoir comparable to that inferred for some radio-emitting TDEs. Whether such a flare is detectable, however, will depend strongly on the density and spatial distribution of the external medium, which are not included in our simulation.

\section{Conclusions}
\label{sec: concl}
Following the results presented in \citet{Martire26} for the TDE of a $0.5M_\odot$ polytropic star by a $10^4M_\odot$ IMBH, we have further analysed the properties of the outflowing material in the simulation, and performed post-processing radiative-transfer calculations to obtain the emerging spectra, single-band light curves and line luminosities. Our continuum post-processing techniques \citep{SS24} agree well with multi-group radiation hydrodynamics \citep{Giron26}. Our results provide additional insight into 
the debated properties of early-time TDE emission, which are especially unconstrained in the IMBH regime.  In particular, we find that:
\begin{enumerate}[(i)]
    \item despite the absence of an accretion disc, an asymmetric wind (i.e., unbound outflow) is launched from the nozzle shock due to its super-Eddington energy dissipation rate;
    
    \item the wind remains strongly anisotropic throughout the simulated evolution, although its geometry becomes progressively more spherically symmetric with time (Fig.~\ref{fig:bern});

    \item as a consequence of this anisotropy, both the physical properties of the wind and the resulting observables depend on the viewing angle, with variations depending not only on the observer's latitude but also with \emph{longitude};

    \item lines of sight oriented towards the pericentre region exhibit properties comparable to those at the poles, being characterised by low densities, low mass loss rates, and high bolometric luminosities, reaching $\approx2$--$3\,L_{\rm Edd}$. In contrast, lines of sight pointed towards the stream display a stronger dependence on latitude: densities and mass-loss rates increase, while the bolometric luminosity decreases, as the line of sight approaches the orbital plane;


    \item the large optical depth of the wind (Fig.~\ref{fig:opac}) supports a novel variant of the classical reprocessing scenario for the early-time emission of TDEs. X-rays initially escape while the wind is absent or optically thin (up to $\approx 1t_{\rm fb}$); as the wind develops and becomes optically thick, the X-ray emission is increasingly reprocessed into UV and optical radiation (Figs.~\ref{fig: mono lc bands}, \ref{fig: Xray lc}).  This sequence of events is conceptually quite similar to classic TDE models where an ionising continuum from a central accretion disc is reprocessed at times and across solid angles where an optically thick reprocessing layer exists \citep{Loeb97, MetzgerStone16, Dai18}, but with the key difference that {\it there is no central accretion disc, and the ionising continuum is completely powered by shocks at small radii}; 

    \item the blackbody temperatures inferred before the peak of the light curve are consistent with those observed in optically selected SMBH TDEs \citep{Hung17, Yao23}. At later times, the system cools more efficiently, resulting in lower temperatures (Fig.~\ref{fig:spectra});

    \item H$\alpha$ and H$\beta$ emission can be detected along lines of sight towards the stream side, with their luminosity evolution delayed relative to that of the bolometric light curve (Fig.~\ref{fig:lines}).

    \item The low peak optical/UV luminosities of the IMBH TDE we have simulated naturally explain the lack of IMBH TDE detections in optical time domain surveys to date \citep{Yao23}, despite their discovery in X-ray surveys \citep{Sazonov21}.  While existing surveys such as ZTF are ultimately too shallow to catch optical/UV flares peaking at $\sim 10^{40.5-41.5}~{\rm erg~s}^{-1}$, near-future surveys by Rubin and {\it ULTRASAT} will likely be able to detect IMBH TDEs out to a horizon of $\sim 790$ and 340 Mpc, respectively.
\end{enumerate}
We have assessed the robustness of our results through convergence tests, finding that variations in the simulation resolution do not affect our qualitative conclusions. Parts of our analysis rely on approximate post-processing calculations to estimate emergent spectra; a more rigorous treatment with Monte Carlo radiative transfer (as in, e.g. \citealt{Roth16, Roth18, Thomsen22, Thomsen26, Mockler26}) would be valuable. As with past RICH simulations \citep{SS24, Martire26, Giron26}, our work is also limited in its ability to resolve small radii inside our gravitational softening radius (i.e. for radii $r<0.6r_{\rm  p}$).

Nevertheless, we have demonstrated that a reprocessing layer can form in IMBH TDEs at early times even without an accretion disc, powered instead by shock dissipation.
The outflow geometry is strongly asymmetric, in contrast to the spherical symmetry assumed in many previous studies \citep{MetzgerStone16, PiroLu20}, which makes 3D simulations necessary to capture the viewing-angle-dependent outflow properties that shape light curves and spectra. 
Despite this angular dependence, the emitted radiation would be detectable by upcoming surveys such as LSST and {\it ULTRASAT} independently on the line of sight, though out to different horizons.
For LSST, the detection horizon ranges from $\approx300$ Mpc when viewed from the stream side to $\approx790$ Mpc from the poles, while for {\it ULTRASAT} it ranges from $\approx90$ to $\approx340$ Mpc, respectively. These near-future optical and UV surveys may therefore uncover a small sample of TDEs from $M_{\rm BH}\sim10^4M_\odot$ IMBHs, whose masses and spins can be more precisely characterised \citep{Wen20, MummeryBalbus20, Wen21, Cao23, Guolo25, Alush25} by X-ray and UV follow-up of their slowly evolving, late-time disc emission.  

Characterisation of nearby IMBHs would be quite valuable on its own, as it may shed light on massive black hole origins \citep{Volonteri10} or the uncertain IMBH occupation fraction \citep{Greene20}.  In the context of TDEs, however, it may also provide a valuable consistency check on mass measurement techniques from late-time disc fitting.  For the foreseeable future, end-to-end simulations of IMBH TDEs will be far cheaper than those of SMBH TDEs, potentially permitting IMBH TDE parameter surveys and radiative transfer post-processing that enable robust determination of black hole properties from {\it early-time} light curves.  Comparison of mass estimates from early-time and late-time emission will be a key cross-check establishing the robustness of TDEs as tools for MBH demographic surveys.

\section*{Acknowledgments} 
PM thanks Yujie He, Pietro Baldini, Biancamaria Sersante, and Manuel Cavieres Carrera  for useful discussions. 
This publication utilises computing time on the Dutch National Computer Facility, Snellius. It is part of the project ``A Library of Disruption Event Simulations: the key to Intermediate Mass Black Hole Discovery'', with file number 022.016, which is funded by the Dutch Research Council (NWO). EMR and PM acknowledge support from the European Research Council (ERC) grant number: 101002511/project acronym: VEGA\_P.  EMR acknowledges that this publication is part of the project ``Flares from disrupted stars unveil the origin of the giant black holes'' with file number  VI.C.232.099 of the research programme VICI, which is financed by NWO. NCS acknowledges support from the Binational Science Foundation (grant No. 2020397), the Israel Science Foundation (Individual Research Grant No. 2414/23), and the European Research Council (ERC) grant number: 1011258072/project acronym: TIDALWAVE. 

\section{Data availability}
The data underlying this article will be shared on reasonable request
to the corresponding author.\\

\emph{Software}: \texttt{RICH} \citep{rich}, numpy \citep{Harris20}, scipy \citep{SciPy20}, h5py, matplotlib \citep{Hunter07}.



\bibliographystyle{mnras}
\bibliography{biblio}

\appendix
\counterwithin{figure}{section}
\renewcommand{\thefigure}{\Alph{section}\arabic{figure}}
\begin{figure*}
    \renewcommand{\thefigure}{A\arabic{figure}}
    \centering
    \includegraphics[width=\linewidth]{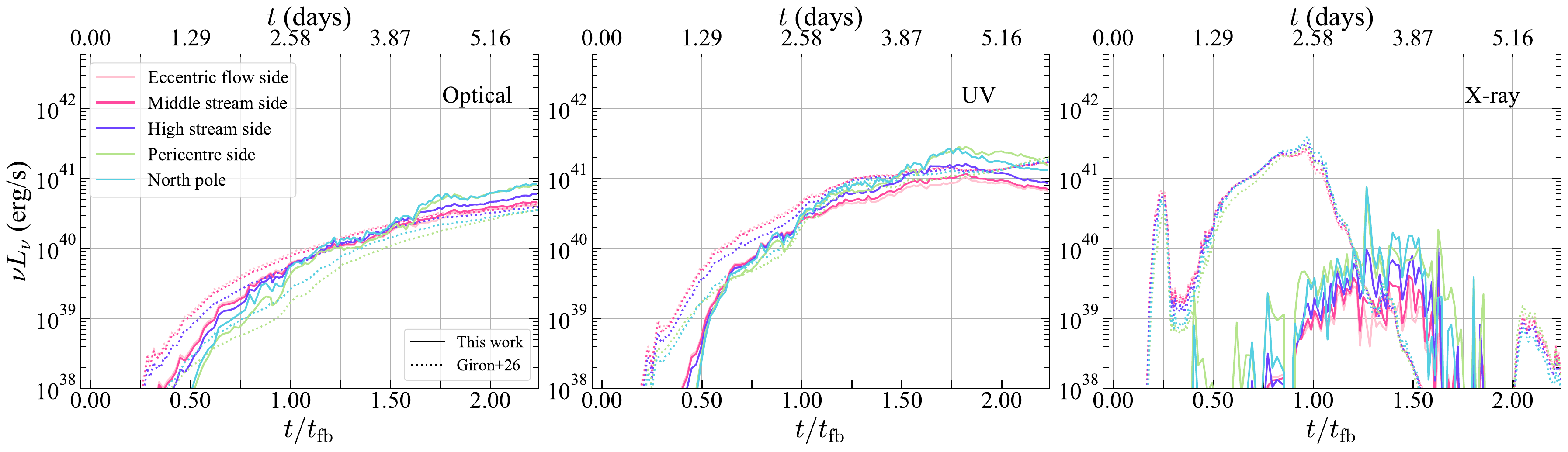}
    \caption{Single-band light curves in optical  (\emph{first panel}), UV (\emph{second panel}) and X-ray (\emph{third panel}) band. Solid and dotted lines represent this work and \citet{Giron26}, respectively. Although the X-ray emission is not well reproduced in the gray approximation, the UV and optical light curves are broadly consistent, differing by a median factor of $\approx1.5-2$ over time in every sector from $1t_{\rm fb}$ onward.}
    \label{fig:MG comp}
\end{figure*}
\section{Opacities}
\label{app: opac}
To find the thermalization radius, we define the effective optical depth as in Eq.\eqref{eq:colorsphere}. Under the assumption of grey radiative transfer, the opacities must be treated as frequency-independent. This can be accomplished by employing mean opacities, such as the Planck or Rosseland averages, which correspond to frequency integrals weighted by the Planck function or its temperature derivative, respectively. Since we are looking at a marginally optically thin regime, it is unclear which choice is more appropriate.
We compute $\tau_{\rm eff}$ numerically as 
\begin{equation}
    \label{eq:num tau}
    \tau_{\rm eff} (r) \equiv \int_{r}^\infty  \sqrt{3\min(\alpha_{\rm abs, P}, \alpha_{\rm Ross})\alpha_{\rm Ross}} \rm{d}r,
\end{equation}
where $\alpha_{\rm abs, P}$ and $\alpha_\text{Ross}$ are, respectively, the Planck and the Rosseland mean absorption coefficient (units of ${\rm cm}^{-1}$) defined as
\begin{equation}
\begin{split} 
\alpha_{\rm abs, P} &= \frac{\int_0^\infty [\alpha_\nu B_\nu(T)] {\rm d}\nu}{\int_0^\infty B_\nu(T) {\rm d}\nu}, \\
    \frac{1}{\alpha_\text{Ross}} &= \frac{\int_0^\infty [(\alpha_\nu+\sigma_\nu)^{-1}\partial B_\nu/\partial T] {\rm d}\nu}{\int_0^\infty[\partial B_\nu/\partial T] {\rm d}\nu},
\end{split}
\end{equation}  
where $\alpha_\nu=\alpha_\nu(T,\rho)$ and $\sigma_\nu=\sigma_\nu(T,\rho)$ are the frequency-dependent absorption and scattering coefficients.  Since our simulation is absorption dominated, scattering is negligible and  $\min(\alpha_{\rm abs, P}, \alpha_{\rm Ross})\approx \alpha_{\rm abs, P}$. If scattering were dominant $\alpha_{\rm Ross}\approx\sigma_{\rm s}\gg \alpha_{\rm abs}$ and $\min(\alpha_{\rm abs, P}, \alpha_{\rm Ross})\approx \alpha_{\rm abs, P}$. In the regime where absorption and scattering contribute comparably,  Eq.\eqref{eq:num tau} would no longer be a reliable implementation of Eq.\eqref{eq:colorsphere}; however, this is not the case for our simulation.

\section{Comparison with multigroup approximation}
\label{app: MG}
\label{app:MG}
In Sec.\ref{sec: lc} we presented single-band light curves computed in post-processing, as described in Sec.\ref{sec:meth}.  Despite the gray approximation adopted in the simulation by \citet{Martire26}, these results remain reliable for the UV and optical bands: their qualitative behaviour is consistent with the multigroup calculations in \citet{Giron26}, with luminosities differing by a median factor of $\approx1.5-2$ over time in every sector from $1t_{\rm fb}$ onward. (see Fig.\ref{fig:MG comp}).
However, because of the treatment of opacities, our method is not applicable to the X-ray band, for which it overestimates the emitted flux.

\section{Parameters for line emission}
\label{app: lines}
We report in Table \ref{tab:line_parameters} the parameter used to compute line emissivities in Fig.\ref{fig:lines}. Values are retrieved from the NIST database\footnote{\url{https://physics.nist.gov/PhysRefData/ASD/lines_form.html}}.

\begin{table}
    \centering
    \caption{Parameters adopted for the calculation of the selected
    hydrogen and helium emission lines from a state at level $u$ to a state at level $l$: \(\chi\) is the ionization
    potential, $\chi_u$ is the excitation energy above the ground state of the
    upper level, \(A\) is the Einstein coefficient, and
    \(\lambda_0\) is the rest wavelength.}
    \label{tab:line_parameters}
    \begin{tabular}{lccc}
        \hline
        Parameter
        & H\(\alpha\)
        & H\(\beta\)
        & He\,\textsc{ii} \(\lambda4686\) \\
        \hline
        Transition \(u\rightarrow l\)
        & \(3\rightarrow2\)
        & \(4\rightarrow2\)
        & \(4\rightarrow3\) \\

        $\chi$ [eV]
        & 13.6
        & 13.6
        & 54.418 \\

        $\chi_{u}$[erg]
        & \(1.9369\times10^{-11}\)
        & \(2.0428\times10^{-11}\)
        & \(8.1738\times10^{-11}\) \\

        $A$ [s\(^{-1}\)]
        & \(6.4651\times10^{7}\)
        & \(8.4193\times10^{6}\)
        & \(2.2076\times10^{8}\) \\

        $\lambda_0$ [cm]
        & \(6.5628\times10^{-5}\)
        & \(4.8613\times10^{-5}\)
        & \(4.6858\times10^{-5}\) \\
        \hline
    \end{tabular}
\end{table}

\bsp	
\label{lastpage}
\end{document}